\documentstyle[amsmath,amssymb,graphicx,wrapfig]{article}

\def\be{\begin{eqnarray}}
\def\ee{\end{eqnarray}}

\hoffset= - 1.0in         

\begin{document}

\thispagestyle{empty}

\baselineskip15pt

\title{{\bf New and Old Results in Resultant Theory}\vspace{0.4cm}}
\author{A.Morozov \footnote{ ITEP, Moscow, Russia;
morozov@itep.ru} \ and 
Sh.Shakirov \thanks{{\small {\it ITEP, Moscow, Russia} and {\it MIPT, Dolgoprudny, Russia}};
shakirov@itep.ru} \date{ }}

\maketitle

\vspace{-4.5cm}

\begin{center}
\hfill ITEP/TH-68/09
\end{center}

\vspace{3.2cm}

\centerline{ABSTRACT}

\vspace{0.6cm}

{\footnotesize
Resultants are getting increasingly important in modern theoretical physics: they appear whenever one deals with non-linear (polynomial) equations, with non-quadratic forms or with non-Gaussian integrals. Being a subject of more than three-hundred-year research, resultants are of course rather well studied: a lot of explicit formulas, beautiful properties and intriguing relationships are known in this field. We present a brief overview of these results, including both recent and already classical. Emphasis is made on explicit formulas for resultants, which could be practically useful in a future physics research. }

\tableofcontents

\section{Introduction}

This paper is devoted to \emph{resultants} -- natural and far-going generalization of \emph{determinants} from conventional linear to generic non-linear systems of algebraic equations

\[
\left\{
\begin{array}{ccc}
f_1(x_1, \ldots, x_n) = 0\\
\\
f_2(x_1, \ldots, x_n) = 0\\
\\
\ldots\\
\\
f_n(x_1, \ldots, x_n) = 0\\
\\
\end{array}
\right.
\]
\smallskip\\
where $f_1, \ldots, f_n$ are homogeneous\footnote{ In this paper we assume homogeneity of the polynomial equations to simplify our considerations; actually, the homogeneity condition can be relaxed and the whole theory can be formulated for non-homogeneous polynomials as well. } polynomials of degrees $r_1, \ldots, r_n$.  This system of $n$ polynomial equations in $n$ variables is over-defined: in general position it does not have non-zero solutions $(x_1, \ldots, x_n)$ at all. A non-vanishing solution exists if and only if a single algebraic constraint is satisfied,

\begin{align}
 R \big\{ f_1, f_2, \ldots, f_n \big\} = 0
\end{align}
\smallskip\\
where $R$ is an irreducible polynomial function called \emph{resultant}, depending on coefficients of the non-linear system under consideration. Clearly, for linear equations resultant is nothing but the determinant -- a familiar object of linear algebra. For non-linear equations, resultant provides a natural generalization of the determinant and therefore can be considered as a central object of \emph{non-linear algebra} \cite{Nolinal}.

Historically, the foundations of resultant theory were laid in the 19th century by Cayley, Sylvester and Bezout \cite{Foundations}. In this period mostly the case of $n = 2$ variables was considered, with some rare exceptions. The theory of resultants for $n > 2$ and closely related objects, namely discriminants and hyperdeterminants, was significantly developed by Gelfand, Kapranov and Zelevinsky in their seminal book \cite{GKZ}.

Let us emphasize, that determinant has a lot of applications in physics. Resultant is expected to have at least as many, though only a few are established yet. This is partly because today's physics is dominated by methods and ideas of linear algebra. These methods are not always successful in dealing with problems, which require an essentially non-linear treatment. We expect that resultants and related algebraic objects will cure this problem and become an important part of the future non-linear physics and for entire string theory program in the sence of \cite{String}. To illustrate this idea, let us briefly describe the relevance of resultants to non-Gaussian integrals -- a topic of undisputable value in physics.

It is commonly accepted, that quantum physics can be formulated in the language of functional integrals. It is hard to question the importance of this approach: since the early works of Wigner, Dirac and Feynman, functional integrals have been a source of new insights and puzzles. Originated in quantum mechanics, they have spread fastly in quantum field theory and finally developed into a widely used tool of modern theoretical physics. From the point of view of functional integration, any quantum theory deals with integrals of a form

\begin{align}
\Big< f(x) \Big> = \int \ d x \ f(x) \ e^{-S(x)}
\label{Correlator}
\end{align}
\smallskip\\
called correlators. The integration variable $x$ belongs to a linear space (which is often infinite-dimensional), and the choice of particular action $S(x)$ specifies a particular theory or a model. In the simplest models, $S(x)$ is merely a quadratic form, which in some particular basis can be written as

\begin{align}
S(x) = \sum\limits_{i,j} S_{ij} x_i x_j
\end{align}
\smallskip\\
If this is the case, the integral (\ref{Correlator}) is called Gaussian. In a sence, Gaussian integrals constitute a foundation of quantum theory: they describe free (non-interacting) particles and fields. These systems are relatively simple, and so are the Gaussian integrals. For example, the simplest two-dimensional Gaussian integral can be evaluated, say, by diagonalisation, and equals

\begin{align}
\int dx \int dy \ e^{-(a x^2 + b xy + c y^2)} = \dfrac{{\rm const}}{\sqrt{ b^2 - 4 ac }}
\end{align}
\smallskip\\
where the expression $b^2 - 4 ac$ can be recognized as a determinant of the quadratic form $a x^2 + b xy + c y^2$. Other Gaussian integrals, with more variables, with non-homogeneous action or with non-trivial insertion $f(\phi)$, are equally easy and similarly related to determinants -- central objects of linear algebra, which describe consistency of linear systems of equations. Remarkably, the study of low-dimensional non-Gaussian integrals of degree three

\begin{align}
S(x) = \sum\limits_{i,j,k} S_{ijk} x_i x_j x_k
\end{align}
\smallskip\\
and higher, reveals their relation to resultants: for example, a two-dimensional non-Gaussian integral equals

\begin{align}
\int dx \int dy \ e^{-(a x^3 + b x^2y + c x y^2 + d y^3)} = \dfrac{{\rm const}}{\sqrt[6]{27 a^2 d^2 - b^2 c^2 - 18 a b c d + 4 a c^3 + 4 b^3 d}}
\end{align}
\smallskip\\
where the expression $27 a^2 d^2 - b^2 c^2 - 18 a b c d + 4 a c^3 + 4 b^3 d$ can be recognized as a \emph{discriminant} of the cubic form $a x^3 + b x^2y + c x y^2 + d y^3$, given by resultant of its derivatives:

\begin{align}
R\Big\{ \ 3 a x^2 + 2 b xy + c y^2, \ b x^2 + 2 c xy + 3 d y^2 \ \Big\} = 27 a^2 d^2 - b^2 c^2 - 18 a b c d + 4 a c^3 + 4 b^3 d
\end{align}
\smallskip\\
In this way, resultants naturally appear in the study of low-dimensional non-Gaussian integrals. This connection between functional integration and non-linear algebra can be traced much further: section 5 of present paper provides a brief summary of this interesting direction of research. See also \cite{IntDisc} for more details.

In this paper we present a selection of topics in resultant theory in a more-or-less explicit manner, which is more convenient for a reader who needs to perform a practical calculation. This also makes our paper accessible to non-experts in the field. Most of statements are explained rather than proved, for some proofs see \cite{GKZ}.

\vspace{-2ex}
\section{Basic Objects}

\vspace{-2ex}
\subsection*{Homogeneous polynomial}

Basic objects of study in non-linear algebra are generic finite-dimensional tensors \cite{Nolinal}. However, especially important in applications are symmetric tensors or, what is the same, homogeneous polynomials of degree $r$ in $n$ variables. In present paper, we restrict consideration to this case. It should be emphasized, that restriction to symmetric tensors is by no means an oversimplification: most of the essential features of non-linear algebra are present at this level. There are two different notations for homogeneous polynomials, which can be useful under different circumstances: tensor notation

$$ S(x_1,x_2,\ldots,x_n) = \sum\limits_{i_1,i_2, \ldots, i_r = 1}^{n} S_{i_1,i_2, \ldots, i_r} x_{i_1} x_{i_2} \ldots x_{i_r} $$
\smallskip\\
and monomial notation

$$ S(x_1,x_2,\ldots,x_n) = \sum\limits_{a_1 + a_2 + \ldots + a_n = r } s_{a_1,a_2 \ldots a_n } x_{1}^{a_1} x_{2}^{a_2} \ldots x_{n}^{a_n} $$
\smallskip\\
To distinguish between these notations, we denote coefficients by capital and small letters, respectively. A homogeneous polynomial of degree $r$ in $n$ variables has $C^{r}_{n+r-1}$ independent coefficients.
\vspace{-2ex}

\subsection*{Resultant}
A system of $n$ polynomial equations, homogeneous of degrees $r_1, \ldots, r_n$ in variables $x_1, \ldots, x_n$

\[
\left\{
\begin{array}{ccc}
f_1(x_1, \ldots, x_n) = 0 \\
\\
f_2(x_1, \ldots, x_n) = 0 \\
\\
\ldots \\
\\
f_n(x_1, \ldots, x_n) = 0 \\
\end{array}
\right.
\]
\smallskip\\
has non-zero solutions, if and only if its coefficients satisfy one algebraic constraint:

$$R_{r_1, \ldots, r_n}\{ f_1, \ldots, f_n \} = 0$$
\smallskip\\
The left hand side of this algebraic constraint is called a resultant of this system of equations. Resultant always exists and is unique up to rescaling. Resultant is a homogeneous polynomial of degree

$$
\deg R_{r_1, \ldots, r_n} = r_1 \ldots r_n \left( \dfrac{1}{r_1} + \ldots + \dfrac{1}{r_n} \right)
$$
\smallskip\\
in the coefficients of the system. Moreover, it is homogeneous in coefficients of each equation $f_i$ separately, and has degree $r_1 \ldots r_{i-1} r_{i+1} \ldots r_n$ in these coefficients. When all equations are linear, i.e, when $r_i = 1$, resultant is equal to the determinant of this system: $R_{11\ldots1}\{f_1,f_2, \ldots, f_n\} = f_1 \wedge f_2 \wedge \ldots \wedge f_n$. Resultant is defined only when the number of homogeneous equations matches the number of variables: when these numbers differ, various more complicated quantities should be used to describe the existence and/or degeneracy of solutions. Some of these quantities are known as higher resultants or subresultants \cite{Higher-1}, some as complanarts \cite{Higher-2}.

\subsection*{Discriminant}

A single homogeneous polynomial $S(x_1, \ldots, x_n)$ of degree $r$ in $n$ variables gives rise to a system of equations

\[
\left\{
\begin{array}{ccc}
\dfrac{\partial S(x_1, \ldots, x_n)}{\partial x_1} = 0 \\
\\
\ldots \\
\\
\dfrac{\partial S(x_1, \ldots, x_n)}{\partial x_n} = 0 \\
\end{array}
\right.
\]
\smallskip\\
which are all homogeneous of degree $r-1$ in variables $x_1, \ldots, x_n$. Resultant of this system is called a discriminant of $S$ and is denoted as $D_{n|r}(S)$. Discriminant is a homogeneous polynomial of degree

$$
\deg D_{n|r}(S) = n(r-1)^{n-1}
$$
\smallskip\\
in the coefficients of $S$, what follows from the analogous formula for the generic resultant. When $S$ is a quadratic form, i.e, when $r = 2$, discriminant is equal to its determinant: $D_{n|2}(S) = \det S$.
\vspace{-1ex}

\subsection*{Invariant}

A Lie group of linear transformations

\[
\left(
\begin{array}{ccc}
x_1 \\
\\
\ldots \\
\\
x_n \\
\end{array}
\right)
\mapsto
\left(
\begin{array}{cccccc}
G_{11} & G_{12} & \ldots & G_{1n} \\
\\
\ldots \\
\\
G_{n1} & G_{12} & \ldots & G_{nn} \\
\end{array}
\right)
\left(
\begin{array}{ccc}
x_1 \\
\\
\ldots \\
\\
x_n \\
\end{array}
\right)
\]
\smallskip\\
with $\det G \neq 0$ is called $GL(n)$. Its subgroup with $\det G = 1$ is called $SL(n)$. In tensor notations, they act

$$x_i \mapsto G_{ij} x_j$$
\smallskip\\
Thus they act on coefficients of homogeneous polynomials $S(x_1, \ldots, x_n)$ of degree $r$ in $n$ variables by the rule

$$S_{i_1, \ldots, i_r} \mapsto G_{i_1 j_1} \ldots G_{i_r j_r} S_{j_1, \ldots, j_r}$$
\smallskip\\
Polynomials $I(S)$ which map to theirselves under $SL(n)$ transformations are called $SL(n)$ invariants of $S$. In particular, discriminant $D_{n|r}(S)$ is a $SL(n)$ invariant of $S$. However, there are many more invariants which have no obvious relation to the discriminant. The number of functionally independent invariants of $S$ is equal to the number of independent coefficients of $S$ minus the dimension of $SL(n)$, that is

$$
\# \mbox{ functionally independent invariants of S} = C^{r}_{n+r-1} - n^2 + 1
$$
\smallskip\\
Therefore, invariants of given degree form a finite-dimensional linear space. All invariants form a ring.

\section{Calculation of resultant}
\subsection{Formulas of Sylvester type }

To find the resultant of $n$ polynomial equations, homogeneous of degree $r$ in variables $x_1, \ldots, x_n$

\[
\left\{
\begin{array}{ccc}
f_1(x_1, \ldots, x_n) = 0 \\
\\
f_2(x_1, \ldots, x_n) = 0 \\
\\
\ldots \\
\\
f_n(x_1, \ldots, x_n) = 0 \\
\end{array}
\right.
\]
\smallskip\\
it is possible to consider another system, multiplied by various monomials:
\begin{align*}
\mbox{ consider } x_1^{s_1} \ldots x_{n}^{s_n} f_i = 0 \mbox{ instead of just } f_i = 0
\end{align*}
In the monomial basis, coefficients of this system can be viewed as a rectangular $n C^{q}_{q+n-1} \times C^{r+q}_{r+q+n-1}$ matrix, where $q = s_1 + \ldots + s_n$ is the multiplier's degree. This matrix is called the (generalized) Sylvester matrix.

\subsubsection{Resultant as a determinant of Sylvester matrix }

In some cases it is possible to choose $q$ in such a way, that Sylvester matrix is square, i.e, $n C^{q}_{q+n-1} = C^{r+q}_{r+q+n-1}$. In these cases, resultant is equal to the determinant of Sylvester matrix. Looking closer at this numeric relation, it is easy to see that the only cases when it is possible are $n = \mbox{any}, r = 1$ and $n = 2, r = \mbox{any}$. In the first case, it suffices to choose $q = 0$ and Sylvester matrix equals the matrix of the system. This is the case of linear algebra. The second case is less trivial: it is necessary to choose $q = r - 1$ to make the Sylvester matrix square. For example, in the case $n = 2, r = 2$ the system of equations has a form

\[
\left\{
\begin{array}{ccc}
f_{11} x_1^2 + f_{12} x_1 x_2 + f_{22} x_2^2 = 0 \\
\\
g_{11} x_1^2 + g_{12} x_1 x_2 + g_{22} x_2^2 = 0 \\
\end{array}
\right.
\]
\smallskip\\
Multiplying the equations of this system by all monomials of degree $q = 1$, that is, by $x_1$ and $x_2$, we obtain

$$
\left\{
\begin{array}{cccc}
f_{11} x_1^3 + f_{12} x_1^2 x_2 + f_{22} x_1 x_2^2 + 0 x_2^3 = 0 \\
\\
0 x_1^3 + f_{11} x_1^2 x_2 + f_{12} x_1 x_2^2 + f_{22} x_2^3 = 0 \\
\\
g_{11} x_1^3 + g_{12} x_1^2 x_2 + g_{22} x_1 x_2^2 + 0 x_2^3 = 0 \\
\\
0 x_1^3 + g_{11} x_1^2 x_2 + g_{12} x_1 x_2^2 + g_{22} x_2^3 = 0 \\
\end{array}\right.
$$
\smallskip\\
These four polynomials are linear combinations of four monomials, i.e, give rise to a square Sylvester matrix:

\begin{align}
R_{2|2}\{f,g\} = \det \left( \begin {array}{cccc}
f_{11}&f_{12}&f_{22}&0\\
\noalign{\medskip}0&f_{11}&f_{12}&f_{22}\\
\noalign{\medskip}g_{11}&g_{12}&g_{22}&0\\
\noalign{\medskip}0&g_{11}&g_{12}&g_{22}\\
\end {array} \right)
\end{align}
\smallskip\\
Formulas, which express resultant as a determinant of some auxillary matrix, are usually called determinantal. In the next case $n = 2, r = 3$ the system of equations has a form

\[
\left\{
\begin{array}{ccc}
f_{111} x_1^3 + f_{112} x_1^2 x_2 + f_{122} x_1 x_2^2 + f_{222} x_2^3 = 0 \\
\\
g_{111} x_1^3 + g_{112} x_1^2 x_2 + g_{122} x_1 x_2^2 + g_{222} x_2^3 = 0 \\
\end{array}
\right.
\]
\smallskip\\
Multiplying the equations of this system by all monomials of degree $q = 2$, that is, by $x_1^2, x_1 x_2$ and $x_2^2$, we get

\begin{align*}
\left\{
\begin{array}{cccc}
f_{111} x_1^5 + f_{112} x_1^4 x_2 + f_{122} x_1^3 x_2^2 + f_{222} x_1^2 x_2^3 + 0 x_1 x_2^4 + 0 x_2^5 = 0 \\
\\
0 x_1^5 + f_{111} x_1^4 x_2 + f_{112} x_1^3 x_2^2 + f_{122} x_1^2 x_2^3 + f_{222} x_1 x_2^4 + 0 x_2^5 = 0 \\
\\
0 x_1^5 + 0 x_1^4 x_2 + f_{111} x_1^3 x_2^2 + f_{112} x_1^2 x_2^3 + f_{122} x_1 x_2^4 + f_{222} x_2^5 = 0 \\
\\
g_{111} x_1^5 + g_{112} x_1^4 x_2 + g_{122} x_1^3 x_2^2 + g_{222} x_1^2 x_2^3 + 0 x_1 x_2^4 + 0 x_2^5 = 0 \\
\\
0 x_1^5 + g_{111} x_1^4 x_2 + g_{112} x_1^3 x_2^2 + g_{122} x_1^2 x_2^3 + g_{222} x_1 x_2^4 + 0 x_2^5 = 0 \\
\\
0 x_1^5 + 0 x_1^4 x_2 + g_{111} x_1^3 x_2^2 + g_{112} x_1^2 x_2^3 + g_{122} x_1 x_2^4 + g_{222} x_2^5 = 0 \\
\end{array}\right.
\end{align*}
\smallskip\\
These four polynomials are linear combinations of four monomials, i.e, give rise to a square Sylvester matrix:

\begin{align}
R_{2|3}\{f,g\} = \det \left(  \begin {array}{cccccc}
f_{111}&f_{112}&f_{122}&f_{222}&0&0\\
\noalign{\medskip}0&f_{111}&f_{112}&f_{122}&f_{222}&0\\
\noalign{\medskip}0&0&f_{111}&f_{112}&f_{122}&f_{222}\\
\noalign{\medskip}g_{111}&g_{112}&g_{122}&g_{222}&0&0\\
\noalign{\medskip}0&g_{111}&g_{112}&g_{122}&g_{222}&0\\
\noalign{\medskip}0&0&g_{111}&g_{112}&g_{122}&g_{222}\\
\end {array} \right)
\end{align}
\smallskip\\
Note, that the matrix has a typical ladder form. The beauty and simplicity made this formula and corresponding resultant $R_{2|r}$ widely known. In our days, Sylvester method is the main apparatus used to calculate two-dimensional resultants and it is the only piece of resultant theory included in undergraduate text books and computer programs such as Mathematica and Maple. However, its generalization to $n > 2$ is not straightforward, since for $n > 2$ and $r \geq 2$ Sylvester matrix is not square.

It turns out that, despite Sylvester matrix in higher dimensions is rectangular, it is still related closely to the resultant. Namely, top-dimensional minors of the Sylvester matrix are always divisible by the resultant, if their size is bigger, than resultant's degree (i.e, for big enough $q$). Various generalizations of the Sylvester method explicitly describe the other, non-resultantal, factor. Therefore, such generalized Sylvester formulas always express the resultant as a ratio of two polynomial quantities. This is certainly a drawback, at least from the computational point of view.

Sylvester's method posesses at least two different generalizations for $n > 2$: one homological, which leads to the theory of so-called Koszul complexes, another more combinatorial, which leads to the construction of so-called Macaulay matrices.

\subsubsection{Resultant as a determinant of Koszul complex }

In construction of the Koszul complex, one complements the commuting variables

$$ x_1, x_2, \ldots, x_n: \ \ x_i x_j - x_j x_i = 0 $$
\smallskip\\
with anti-commuting variables

$$ \theta_1, \theta_2, \ldots, \theta_n: \ \ \theta_i \theta_j + \theta_j \theta_i = 0 $$
\smallskip\\
and considers polynomials, depending both on $x_1, \ldots, x_n$ and $\theta_1, \ldots, \theta_n$.
Denote through $\Omega(p,q) $ the space of such polynomials of degree $p$ in $x$-variables and degree $q$ in $\theta$-variables. The degree $q$ can not be greater than $n$, since $\theta$ are anti-commuting variables. Dimensions of these spaces are

\begin{align}
\mbox{ dim } \Omega(p,q) = C^{p}_{p+n-1} C^{q}_{n}
\label{Dims}
\end{align}
\smallskip\\
Koszul differential, built from $f$, is a linear operator which acts on these spaces by the rule

\begin{align}
 \hat d = f_1 (x_1, \ldots, x_n) \dfrac{\partial}{\partial \theta_1} + \ldots + f_n (x_1, \ldots, x_n) \dfrac{\partial}{\partial \theta_n}
\label{Differential}
\end{align}
\smallskip\\
and is automatically nilpotent:

$$ \hat d\hat d = f_j(x)f_k(x)\frac{\partial}{\partial\theta_j}\frac{\partial}{\partial\theta_k} = f_j(x)f_k(x) \left( \frac{\partial}{\partial\theta_j}\frac{\partial}{\partial\theta_k} + \frac{\partial}{\partial\theta_k}\frac{\partial}{\partial\theta_j} \right) = 0$$
\smallskip\\
It sends

$$ \hat d: \ \Omega(p,q) \rightarrow \Omega(p + r, q - 1) $$
\smallskip\\
giving rise to the following Koszul complex:

$$\Omega(p,q)\stackrel{\hat d}{\rightarrow}\Omega(p+r,q-1)\stackrel{\hat d}{\rightarrow}\Omega(p+2r,q-2)\stackrel{\hat d}{\rightarrow}...\stackrel{\hat d}{\rightarrow}\Omega(0,R)$$
\smallskip\\
Thus, for one and the same operator $\hat d$ there are many different Koszul complexes,
depending on a single integer parameter $R$ -- the $x$-degree of the rightmost space.
Using the formula (\ref{Dims}) for dimensions, it is easy to write down all of them.
For example, the tower of Koszul complexes for the case $3|2$ is

\begin{center}
$\boxed{ 3 \vert 2 }$ \ \
\begin{tabular}{cccc}
R & Spaces & Dimensions & Euler characteristic \\
\hline
$0$ & $\Omega(0,0)$ & $1$ & $1$ \\
$1$ & $\Omega(1,0)$ & $3$ & $3$ \\
$2$ & $\Omega(0,1) \rightarrow \Omega(2,0)$ & $3 \rightarrow 6$ & $3$ \\
$3$ & $\Omega(1,1) \rightarrow \Omega(3,0)$ & $9 \rightarrow 10$ & $1$ \\
$4$ & $\Omega(0,2) \rightarrow \Omega(2,1) \rightarrow \Omega(4,0)$ & $3 \rightarrow 18 \rightarrow 15$ & $0$ \\
$5$ & $\Omega(1,2) \rightarrow \Omega(3,1) \rightarrow \Omega(5,0)$ & $9 \rightarrow 30 \rightarrow 21$ & $0$ \\
$6$ & $\Omega(0,3) \rightarrow \Omega(2,2) \rightarrow \Omega(4,1) \rightarrow \Omega(6,0)$ & $1 \rightarrow 18 \rightarrow 45 \rightarrow 28$ & $0$ \\
$7$ & $\Omega(1,3) \rightarrow \Omega(3,2) \rightarrow \Omega(5,1) \rightarrow \Omega(7,0)$ & $3 \rightarrow 30 \rightarrow 63 \rightarrow 36$ & $0$ \\
$\ldots$ & $\ldots$ & $\ldots$ & $\ldots$
\end{tabular}
\smallskip\\
\end{center}
where Euler characteristic is the alternating combination of dimensions. The rightmost differential

\begin{align*}
 \hat d: \ \Omega(R - r, 1) \stackrel{\hat d}{\rightarrow} \Omega(R, 0)
\end{align*}
\smallskip\\
acts on the corresponding linear spaces as

\begin{align*}
 \Big( x_1^{i_1} x_2^{i_2} \ldots x_n^{i_n} \Big) \cdot \theta_j \mapsto \Big( x_1^{i_1} x_2^{i_2} \ldots x_n^{i_n} \Big) \cdot f_j\big( x_1, \ldots, x_n \big)
\end{align*}
\smallskip\\
and is represented by $n C^{R - r}_{R - r + n - 1} \times C^{R}_{R + n - 1} $ matrix. It is easy to see, that this is exactly the rectangular Sylvester matrix. Therefore, Koszul complex in a sence complements the Sylvester matrix with many other matrices, which carry additional information. Resultant is equal to certain characteristic of this complex

\begin{equation}
\addtolength{\fboxsep}{5pt}
\boxed{
\begin{gathered}
R_{n|r}\{ f_1, \ldots, f_n \} = \mbox{ DET } (d_1, d_2, \ldots, d_{p-1} )
\end{gathered}
}\label{MainEquality}
\end{equation}
\smallskip\\
which is also called determinant, not to be confused with ordinary determinant of a linear map. Here $d_i$ are the rectangular matrices, representing particular terms of the complex. The "determinantal" terminology was introduced by Cayley, who first studied $\mbox{ DET } (d_1, d_2, \ldots, d_{p-1} )$, and is rather unusual for modern literature, where $\mbox{ DET } (d_1, d_2, \ldots, d_{p-1} )$ is called a (Reidemeister) torsion of the complex $d_1, \ldots, d_{p-1}$. Since the quantity itself is a rather natural generalisation of determinants from linear maps to complexes, we prefer to use the historical name "determinant of a complex".

Calculation of determinants of complexes is described in some detail in \cite{GKZ, Koszul}. Given a complex

\begin{align}
L_1\stackrel{d_1}{\longrightarrow} L_2\stackrel{d_2}{\longrightarrow} ...\stackrel{d_{p-1}}{\longrightarrow}L_p
\label{ComplexForDet}
\end{align}
\smallskip\\
with vanishing Euler characteristic

\begin{align*}
\chi = \mbox{dim }L_1 - \mbox{dim }L_2 + \mbox{dim }L_3 - \ldots + (-1)^p \ \mbox{dim }L_p = 0
\end{align*}
\smallskip\\
one needs to select a basis in each linear space $L_{i}$ and label the basis vectors by elements of a set

\begin{align*}
\big\{ 1,2, \ldots, l_i \big\}
\end{align*}
\smallskip\\
After the basis is chosen, the complex is represented by a sequence of rectangular matrices. The $l_i \times l_{i+1}$ matrix $d_i$ has $l_i$ rows and $l_{i+1}$ coloums. To calculate the determinant of this complex, one needs to select arbitrary subsets $ \sigma_i \subset \big\{ 1,2, \ldots, l_i \big\}$ which consist of

\begin{align*}
 \big| \sigma_i \big| = \sum\limits_{k = 1}^{i-1} (-1)^{i + k + 1} l_k
\end{align*}
elements, i.e,

\begin{align*}
\big| \sigma_1 \big| = 0, \ \ \ \ \big| \sigma_2 \big| = l_1, \ \ \ \ \big| \sigma_3 \big| = l_2 - l_1, \ \ \ \ \big| \sigma_4 \big| = l_3 - l_2 + l_1, \ \ \ \ \ldots
\end{align*}
\smallskip\\
and define conjugate subsets $ \widetilde{ \sigma}_i = \big\{ 1,2, \ldots, l_i \big\} / \sigma_i $ which consist of

\begin{align*}
\big| \widetilde{ \sigma}_i \big| = l_i - \big| \sigma_i \big| = \big| \sigma_{i+1} \big|
\end{align*}
\smallskip\\
elements. Finally, one needs to calculate minors

\begin{align*}
M_i = \mbox{ the minor of } d_i\mbox{, which occupies rows } \widetilde{ \sigma}_i \mbox{ and coloumns } \sigma_{i+1} \end{align*}
\smallskip\\
The determinant of complex is then given by

\begin{equation}
\addtolength{\fboxsep}{5pt}
\boxed{
\begin{gathered}
\mbox{ DET } (d_1, d_2, \ldots, d_{p-1} ) = \prod\limits_{i = 1}^{p-1} \big( M_{i} \big)^{(-1)^{p + i + 1}} = \dfrac{M_{p-1} M_{p - 3} \ldots }{M_{p - 2} M_{p - 4} \ldots }
\end{gathered}
}\label{DetComplexRatio}
\end{equation}
\smallskip\\
In fact, this answer does not depend on the choice of $\sigma_i$ \cite{Koszul} and thus gives a practical way to calculate resultants of polynomial systems. Given a system of equations, one straightforwardly constructs the differential (\ref{Differential}) and the linear spaces (\ref{Dims}), calculates the matrices and their minors according to the above procedure and, finally, uses the formula (\ref{DetComplexRatio}) to find the resultant.

The answer, obtained in this way, is excessively complicated in the following two senses: (i) the ratio of various determinants at the r.h.s. is actually reduced to a polynomial, but this cancelation is not explicit in (\ref{MainEquality}); (ii) the r.h.s. is actually independent of the choice of Koszul complex, but this independence is not explicit in (\ref{MainEquality}). Still, Koszul complex provides an explicit tool of calculation of resultants, which is reasonably useful for particular low $n$ and $r$.

\pagebreak

To illustrate the Koszul complex method, let us consider in detail the case $n = 3, r = 2$, a system of three quadratic equations in three unknowns. Let us denote their coefficients by $a,b,c$ instead of usual $f,g,h$:

$$
\left\{ \begin{array}{l}
f_1(x_1,x_2,x_3) = a_{11} x_1^2 + a_{12} x_1 x_2 + a_{13} x_1 x_3 + a_{22} x_2^2 + a_{23} x_2 x_3 + a_{33} x_{3}^2 \\
\\
f_2(x_1,x_2,x_3) = b_{11} x_1^2 + b_{12} x_1 x_2 + b_{13} x_1 x_3 + b_{22} x_2^2 + b_{23} x_2 x_3 + b_{33} x_{3}^2 \\
\\
f_3(x_1,x_2,x_3) = c_{11} x_1^2 + c_{12} x_1 x_2 + c_{13} x_1 x_3 + c_{22} x_2^2 + c_{23} x_2 x_3 + c_{33} x_{3}^2 \\
\end{array}\right.
$$
\smallskip\\
By definition, Koszul differential is given by

\begin{align*}
\hat d =
f_1\frac{\partial}{\partial\theta_1}+f_2\frac{\partial}{\partial\theta_2}+f_3\frac{\partial}{\partial\theta_3}
\end{align*}
\smallskip\\
Let us describe in detail the simplest case $R = 4$ and next-to-the-simplest case $R = 5$.
The complex with $R = 4$, is the first determinant-possessing complex (i.e. has a vanishing Euler characteristics):

\begin{align*}
\Omega(0,2)\rightarrow\Omega(2,1)\rightarrow\Omega(4,0)
\end{align*}
\smallskip\\
with dimensions $ 3 \rightarrow 18 \rightarrow 15$. If we select a basis in $\Omega(0,2)$ as

\begin{align*}
\{ \theta_2 \theta_3, \theta_1 \theta_3, \theta_1 \theta_2 \}
\end{align*}
\smallskip\\
a basis in $\Omega(2,1)$ as

\begin{align*}
\{ x_{1}^2 \theta_{3}, x_{1}^2 \theta_{2}, x_{1}^2 \theta_{1}, x_{1} x_{2} \theta_{3}, x_{1} x_{2} \theta_{2}, x_{1} x_{2} \theta_{1}, x_{1} x_{3} \theta_{3}, x_{1} x_{3} \theta_{2},
\end{align*}

\begin{align*}
x_{1} x_{3} \theta_{1}, x_{2}^2 \theta_{3}, x_{2}^2 \theta_{2}, x_{2}^2 \theta_{1}, x_{2} x_{3} \theta_{3}, x_{2} x_{3} \theta_{2}, x_{2} x_{3} \theta_{1}, x_{3}^2 \theta_{3}, x_{3}^2 \theta_{2}, x_{3}^2 \theta_{1} \}
\end{align*}
\smallskip\\
and, finally, a basis in $\Omega(4,0)$ as

\begin{align*}
\{ x_{1}^4, x_{1}^3 x_{2}, x_{1}^3 x_{3}, x_{1}^2 x_{2}^2, x_{1}^2 x_{2} x_{3}, x_{1}^2 x_{3}^2, x_{1} x_{2}^3, x_{1} x_{2}^2 x_{3}, x_{1} x_{2} x_{3}^2, x_{1} x_{3}^3, x_{2}^4, x_{2}^3 x_{3}, x_{2}^2 x_{3}^2, x_{2} x_{3}^3, x_{3}^4 \} \end{align*}
\smallskip\\
then Koszul differential is represented by a pair of matrices: one $3 \times 18$

\begin{center}
\includegraphics[width=350pt]{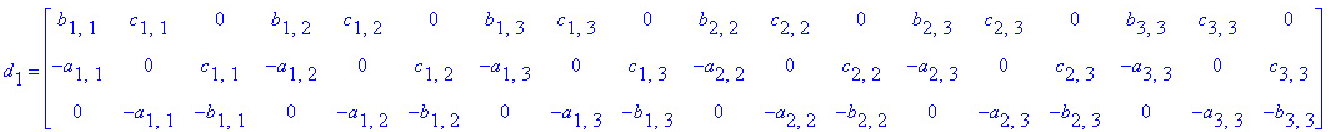}
\end{center}
and another $18 \times 15$

\begin{center}
\includegraphics[width=350pt]{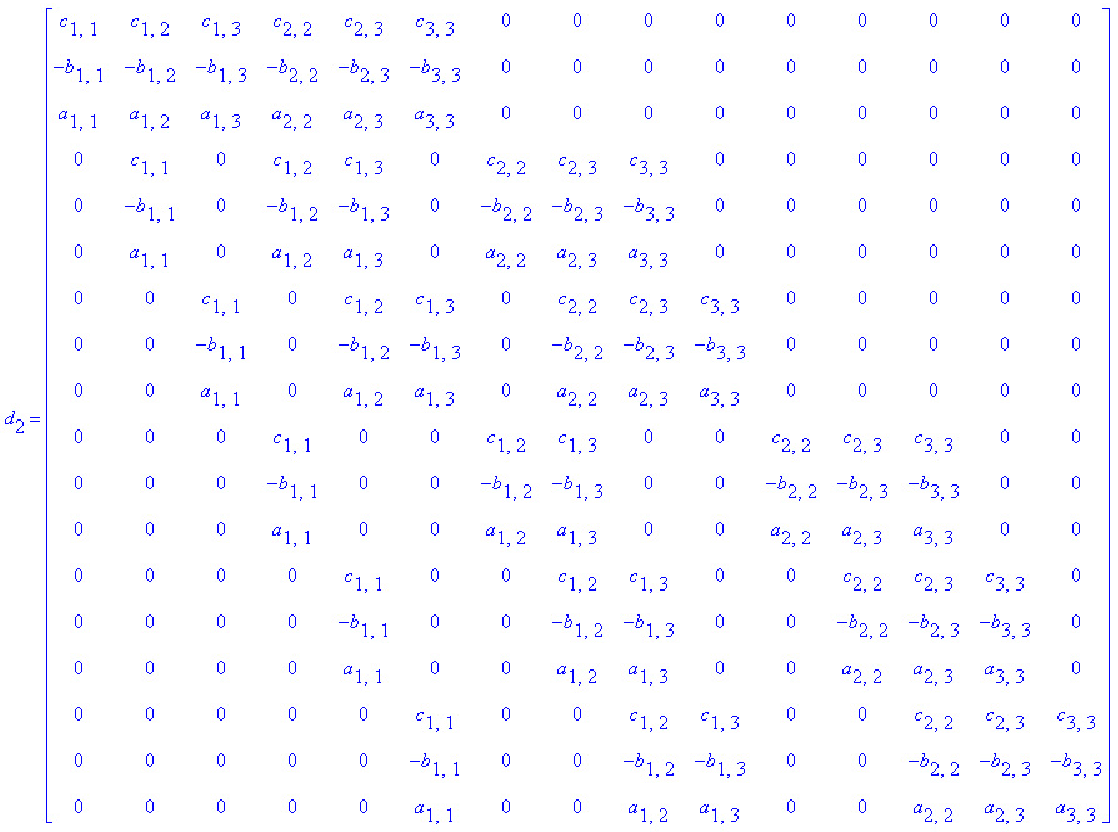}
\end{center}
By selecting some 3 columns in $\hat d_1$ and complementary 15 rows in
$\hat d_2$, we obtain the desired resultant

\begin{center}
\includegraphics[width=300pt]{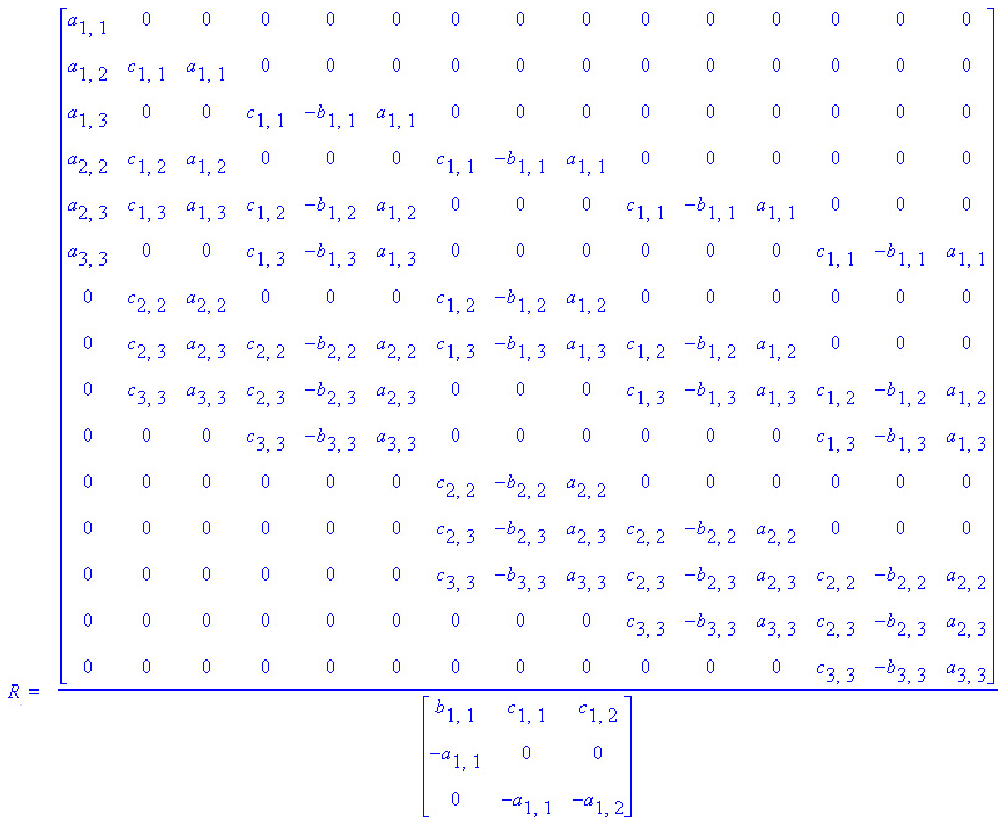}
\end{center}
This particular formula corresponds to the choice of columns 1, 2 and 5
in the first matrix, but the resultant, of course, does not
depend on this choice: any other three columns will do. One only needs
to care that determinant in the denominator does not vanish. If it is
non-zero, then the upper minor is divisible by the lower minor, and
the fraction is equal (up to sign) to the resultant.

\pagebreak

Alternative complex is the second complex with vanishing $\chi$, that corresponds to $R = 5$: $$ \Omega(1,2)\rightarrow\Omega(3,1)\rightarrow\Omega(5,0)$$ It has dimensions $ 9 \rightarrow 30 \rightarrow 21$. If we select a basis in $\Omega(1,2)$ as
$$ \{ x_{1} \theta_{2} \theta_{3}, x_{1} \theta_{1} \theta_{3}, x_{1} \theta_{1} \theta_{2}, x_{2} \theta_{2} \theta_{3}, x_{2} \theta_{1} \theta_{3}, x_{2} \theta_{1} \theta_{2}, x_{3} \theta_{2} \theta_{3}, x_{3} \theta_{1} \theta_{3}, x_{3} \theta_{1} \theta_{2} \} $$
a basis in $\Omega(3,1)$ as
$$ \{ x_{1}^3 \theta_{3}, x_{1}^3 \theta_{2}, x_{1}^3 \theta_{1}, x_{1}^2 x_{2} \theta_{3}, x_{1}^2 x_{2} \theta_{2}, x_{1}^2 x_{2} \theta_{1}, x_{1}^2 x_{3} \theta_{3}, x_{1}^2 x_{3} \theta_{2}, x_{1}^2 x_{3} \theta_{1}, $$
$$ x_{1} x_{2}^2 \theta_{3}, x_{1} x_{2}^2 \theta_{2}, x_{1} x_{2}^2 \theta_{1}, x_{1} x_{2} x_{3} \theta_{3}, x_{1} x_{2} x_{3} \theta_{2}, x_{1} x_{2} x_{3} \theta_{1}, x_{1} x_{3}^2 \theta_{3}, x_{1} x_{3}^2 \theta_{2}, x_{1} x_{3}^2 \theta_{1}, x_{2}^3 \theta_{3}, x_{2}^3 \theta_{2}, x_{2}^3 \theta_{1}, $$ $$x_{2}^2 x_{3} \theta_{3}, x_{2}^2 x_{3} \theta_{2}, x_{2}^2 x_{3} \theta_{1}, x_{2} x_{3}^2 \theta_{3}, x_{2} x_{3}^2 \theta_{2}, x_{2} x_{3}^2 \theta_{1}, x_{3}^3 \theta_{3}, x_{3}^3 \theta_{2}, x_{3}^3 \theta_{1} $$
$$ x_{1} x_{3} \theta_{1}, x_{2}^2 \theta_{3}, x_{2}^2 \theta_{2}, x_{2}^2 \theta_{1}, x_{2} x_{3} \theta_{3}, x_{2} x_{3} \theta_{2}, x_{2} x_{3} \theta_{1}, x_{3}^2 \theta_{3}, x_{3}^2 \theta_{2}, x_{3}^2 \theta_{1} \} $$
and, finally, a basis in $\Omega(5,0)$ as
$$ \{ x_{1}^5, x_{1}^4 x_{2}, x_{1}^4 x_{3}, x_{1}^3 x_{2}^2, x_{1}^3 x_{2} x_{3}, x_{1}^3 x_{3}^2, x_{1}^2 x_{2}^3, x_{1}^2 x_{2}^2 x_{3}, x_{1}^2 x_{2} x_{3}^2, x_{1}^2 x_{3}^3, $$ $$ x_{1} x_{2}^4, x_{1} x_{2}^3 x_{3}, x_{1} x_{2}^2 x_{3}^2, x_{1} x_{2} x_{3}^3, x_{1} x_{3}^4, x_{2}^5, x_{2}^4 x_{3}, x_{2}^3 x_{3}^2, x_{2}^2 x_{3}^3, x_{2} x_{3}^4, x_{3}^5 \} $$
then Koszul differential is represented by a pair of matrices: one $9 \times 30$
\begin{center}
\includegraphics[width=400pt]{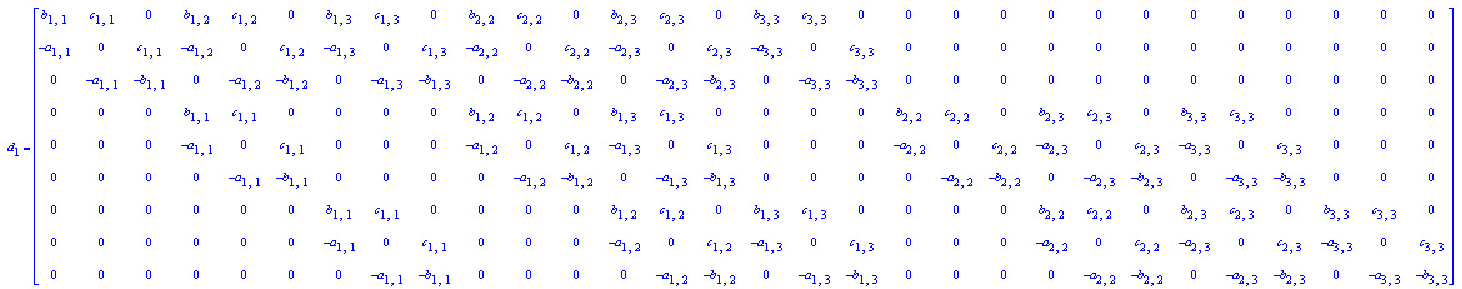}
\end{center}
and another $30 \times 21$
\begin{center}
\includegraphics[width=250pt]{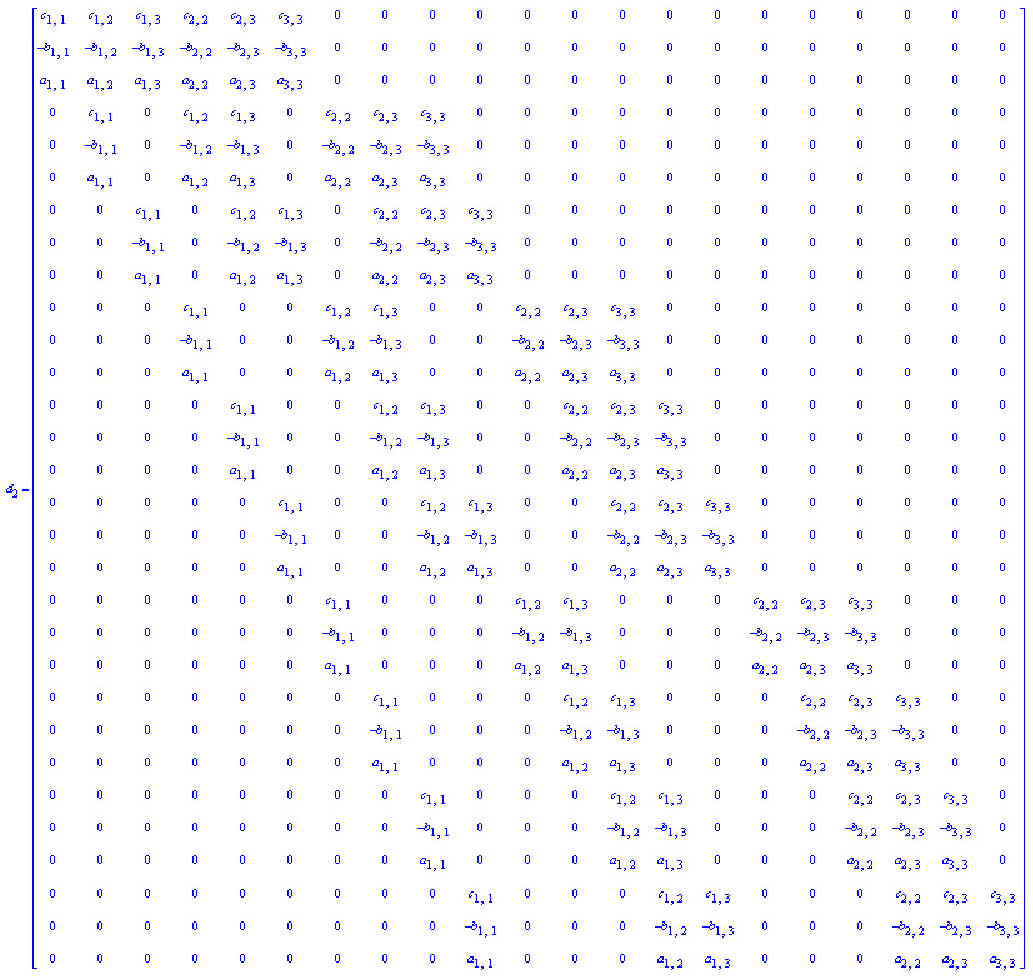}
\end{center}
By selecting some 9 columns in $\hat d_1$ and complementary 21 rows in
$\hat d_2$, we obtain the desired resultant
\begin{center}
\includegraphics[width=450pt]{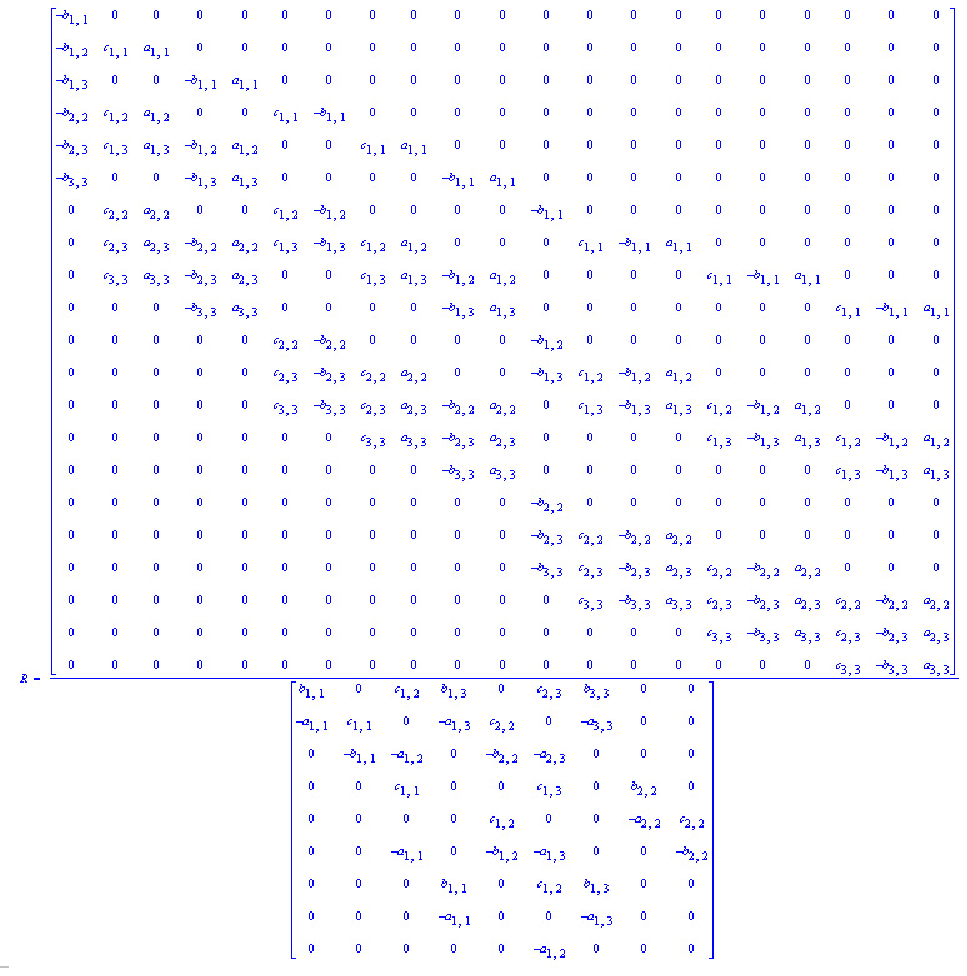}
\end{center}
This particular formula corresponds to the choice of columns
$1,3,5,7,12,14,16,19$ and $21$ in the first matrix. Of
course, any other choice will give the the same (up to a sign) result,
if only the minor in denominator is non-zero. Direct calculation
(division of one minor by another) should demonstrate, that the two complexes $ 3
\rightarrow 18 \rightarrow 15$ and $ 9 \rightarrow 30 \rightarrow
21$ give one and the same expression for the resultant.
It should also be reproduced by determinants of all other complexes
from the $3|2$ list with $R > 3$.

\subsubsection{Macaulay's formula }

The second, rather combinatorial, generalization of the Sylvester method is due to Macaulay. This method allows to express the resultant as a ratio of just two determinants (for Koszul complexes (\ref{DetComplexRatio}) both the numerator and denominator are products of several determinants). For this reason, Macaulay formula is computationally more economical, than eq. (\ref{DetComplexRatio}). We do not go into details here, they can be found in \cite{Macaulay}.

\pagebreak

\subsection{Formulas of Bezout type }

Consider a system of $n$ polynomial equations, homogeneous of degree $r$ in variables $x_1, \ldots, x_n$:

\[
\left\{
\begin{array}{ccc}
f_1(x_1, \ldots, x_n) = 0 \\
\\
f_2(x_1, \ldots, x_n) = 0 \\
\\
\ldots \\
\\
f_n(x_1, \ldots, x_n) = 0 \\
\end{array}
\right.
\]
\smallskip\\
where

$$ f_i(x_1, \ldots, x_n) = \sum\limits_{a_1 + a_2 + \ldots + a_n = r } f_{i | a_1 a_2 \ldots a_n } x_{1}^{a_1} x_{2}^{a_2} \ldots x_{n}^{a_n} $$
\smallskip\\
In the monomial basis, coefficients of this system can be viewed as forming an $n \times C^{r}_{n+r-1}$ rectangular matrix. The set of all top-dimensional minors of this matrix is naturally a tensor with $n$ indices:

$$ M_{i_1 \ldots i_n} = n \times n \mbox{ minor of } f \mbox{, situated in coloumns } i_1, \ldots, i_n $$
\smallskip\\
Resultant $R\{f_1, \ldots, f_n\}$ is expected to be a homogeneous polynomial of variables $M_{i_1 \ldots i_n}$ of degree $r^{n-1}$. Such expressions are called formulas of Bezout type. The simplest example of Bezout formulas is the case $n = 2, r = 1$, where the system of equations is just a simple linear system:

\[
\left\{
\begin{array}{ccc}
f_1 x_1 + f_2 x_2 = 0 \\
\\
g_1 x_1 + g_2 x_2 = 0 \\
\end{array}
\right.
\]
\smallskip\\
and the only top-dimensional minor is

$$M_{12} = f_1 g_2 - f_2 g_1$$
\smallskip\\
In this case, resultant is just equal to this minor:

$$R_{2|1}\{f, g\} = M_{12}$$
\smallskip\\
In the next-to-simplest case $n = 2, r = 2$ the system of equations has a form

\[
\left\{
\begin{array}{ccc}
f_{11} x_1^2 + f_{12} x_1 x_2 + f_{22} x_2^2 = 0 \\
\\
g_{11} x_1^2 + g_{12} x_1 x_2 + g_{22} x_2^2 = 0 \\
\end{array}
\right.
\]
\smallskip\\
and there are three top-dimensional minors:

\[
\begin{array}{ccc}
M_{12} = f_{11} g_{12} - f_{12} g_{11}\\
\\
M_{13} = f_{11} g_{22} - f_{22} g_{11}\\
\\
M_{23} = f_{12} g_{22} - f_{22} g_{12}\\
\end{array}
\]
\smallskip\\
In this case, resultant is a homogeneous polynomial of degree $2$ in minors , which is moreover a determinant

$$R_{2|2}\{f, g\} = M_{12} M_{23} - M_{13}^2 = \det \left( \begin{array}{cc} M_{12} & M_{13} \\ \\ M_{13} & M_{23} \end{array} \right)$$
\smallskip\\
For $n = 2, r = 3$ the system of equations has a form

\[
\left\{
\begin{array}{ccc}
f_{111} x_1^3 + f_{112} x_1^2 x_2 + f_{122} x_1 x_2^2 + f_{222} x_2^3 = 0 \\
\\
g_{111} x_1^3 + g_{112} x_1^2 x_2 + g_{122} x_1 x_2^2 + g_{222} x_2^3 = 0 \\
\end{array}
\right.
\]
\smallskip\\
and there are six top-dimensional minors:

\[
\begin{array}{ccc}
M_{12} = f_{111} g_{112} - f_{112} g_{111} & M_{13} = f_{111} g_{122} - f_{122} g_{111}\\
\\
M_{14} = f_{111} g_{222} - f_{222} g_{111} & M_{23} = f_{112} g_{122} - f_{122} g_{112}\\
\\
M_{24} = f_{112} g_{222} - f_{222} g_{112} & M_{34} = f_{122} g_{222} - f_{222} g_{122}\\
\end{array}
\]
\smallskip\\
In this case, resultant is a homogeneous polynomial of degree $3$ in minors, which is again a determinant

\begin{align}
R_{2|3}\{f, g\} = M_{1 2} M_{3 4} M_{1 4}+M_{1 2} M_{3 4} M_{2 3}-M_{1 2} M_{2 4}^2-M_{1 3}^2 M_{3 4}+2 M_{1 3} M_{1 4} M_{2 4}-M_{1 4}^3-M_{1 4}^2 M_{2 3}
\end{align}

\begin{align}
R_{2|3}\{f, g\} = \det \left( \begin{array}{ccc} M_{12} & M_{13} & M_{14} \\ \\ M_{13} & M_{23} + M_{14} & M_{23} \\ \\ M_{14} & M_{24} & M_{34} \\ \\ \end{array} \right)
\end{align}
\smallskip\\
Note, that six variables $M_{ij}$ are not independent: they are subject to one constraint

\begin{align}
M_{12} M_{34} - M_{13} M_{24} + M_{14} M_{23} = 0
\end{align}
\smallskip\\
which is well-known as Plucker relation. For this reason, the polynomial expression is actually not unique: there are many different polynomial expressions for resultant through minors, whose difference is proportional to the Plucker relation. Generally (for higher $n$ and $r$) there are several Plucker relations, not a single one. Variables $M_{i_1 \ldots i_n}$ are sometimes called Plucker coordinates. Bezout formulas, which express resultant through Plucker coordinates, exist in higher dimensions: for example, for $n = 3, r = 3$ we have

\[
\left\{ \begin{array}{c}
f(x_1, x_2, x_3) = f_{11} x_1^2 + f_{12} x_1 x_2 + f_{13} x_1 x_3 + f_{22} x_2^2 + f_{23} x_2 x_3 + f_{33} x_3^2 = 0 \\
\noalign{\medskip}g(x_1, x_2, x_3) = g_{11} x_1^2 + g_{12} x_1 x_2 + g_{13} x_1 x_3 + g_{22} x_2^2 + g_{23} x_2 x_3 + g_{33} x_3^2 = 0 \\
\noalign{\medskip}h(x_1, x_2, x_3) = h_{11} x_1^2 + h_{12} x_1 x_2 + h_{13} x_1 x_3 + h_{22} x_2^2 + h_{23} x_2 x_3 + h_{33} x_3^2 = 0 \\
\end{array} \right.
\]
\smallskip\\
and there are 20 top-dimensional minors:

\[
\begin{array}{c}
M_{123} = f_{11} g_{12} h_{13}-f_{11} g_{13} h_{12}-g_{11} f_{12} h_{13}+g_{11} f_{13} h_{12}+h_{11} f_{12} g_{13}-h_{11} f_{13} g_{12}\\
\\
M_{124} = f_{11} g_{12} h_{22}-f_{11} g_{22} h_{12}-g_{11} f_{12} h_{22}+g_{11} f_{22} h_{12}+h_{11} f_{12} g_{22}-h_{11} f_{22} g_{12}\\
\\
M_{125} = f_{11} g_{12} h_{23}-f_{11} g_{23} h_{12}-g_{11} f_{12} h_{23}+g_{11} f_{23} h_{12}+h_{11} f_{12} g_{23}-h_{11} f_{23} g_{12}\\
\\
M_{126} = f_{11} g_{12} h_{33}-f_{11} g_{33} h_{12}-g_{11} f_{12} h_{33}+g_{11} f_{33} h_{12}+h_{11} f_{12} g_{33}-h_{11} f_{33} g_{12}\\
\\
M_{134} = f_{11} g_{13} h_{22}-f_{11} g_{22} h_{13}-g_{11} f_{13} h_{22}+g_{11} f_{22} h_{13}+h_{11} f_{13} g_{22}-h_{11} f_{22} g_{13}\\
\\
M_{135} = f_{11} g_{13} h_{23}-f_{11} g_{23} h_{13}-g_{11} f_{13} h_{23}+g_{11} f_{23} h_{13}+h_{11} f_{13} g_{23}-h_{11} f_{23} g_{13}\\
\\
M_{136} = f_{11} g_{13} h_{33}-f_{11} g_{33} h_{13}-g_{11} f_{13} h_{33}+g_{11} f_{33} h_{13}+h_{11} f_{13} g_{33}-h_{11} f_{33} g_{13}\\
\\
M_{145} = f_{11} g_{22} h_{23}-f_{11} g_{23} h_{22}-g_{11} f_{22} h_{23}+g_{11} f_{23} h_{22}+h_{11} f_{22} g_{23}-h_{11} f_{23} g_{22}\\
\\
M_{146} = f_{11} g_{22} h_{33}-f_{11} g_{33} h_{22}-g_{11} f_{22} h_{33}+g_{11} f_{33} h_{22}+h_{11} f_{22} g_{33}-h_{11} f_{33} g_{22}\\
\\
M_{156} = f_{11} g_{23} h_{33}-f_{11} g_{33} h_{23}-g_{11} f_{23} h_{33}+g_{11} f_{33} h_{23}+h_{11} f_{23} g_{33}-h_{11} f_{33} g_{23}\\
\\
M_{234} = f_{12} g_{13} h_{22}-f_{12} g_{22} h_{13}-g_{12} f_{13} h_{22}+g_{12} f_{22} h_{13}+h_{12} f_{13} g_{22}-h_{12} f_{22} g_{13}\\
\\
M_{235} = f_{12} g_{13} h_{23}-f_{12} g_{23} h_{13}-g_{12} f_{13} h_{23}+g_{12} f_{23} h_{13}+h_{12} f_{13} g_{23}-h_{12} f_{23} g_{13}
\end{array}
\]

\[
\begin{array}{ccc}
M_{236} = f_{12} g_{13} h_{33}-f_{12} g_{33} h_{13}-g_{12} f_{13} h_{33}+g_{12} f_{33} h_{13}+h_{12} f_{13} g_{33}-h_{12} f_{33} g_{13}\\
\\
M_{245} = f_{12} g_{22} h_{23}-f_{12} g_{23} h_{22}-g_{12} f_{22} h_{23}+g_{12} f_{23} h_{22}+h_{12} f_{22} g_{23}-h_{12} f_{23} g_{22}\\
\\
M_{246} = f_{12} g_{22} h_{33}-f_{12} g_{33} h_{22}-g_{12} f_{22} h_{33}+g_{12} f_{33} h_{22}+h_{12} f_{22} g_{33}-h_{12} f_{33} g_{22}\\
\\
M_{256} = f_{12} g_{23} h_{33}-f_{12} g_{33} h_{23}-g_{12} f_{23} h_{33}+g_{12} f_{33} h_{23}+h_{12} f_{23} g_{33}-h_{12} f_{33} g_{23}\\
\\
M_{345} = f_{13} g_{22} h_{23}-f_{13} g_{23} h_{22}-g_{13} f_{22} h_{23}+g_{13} f_{23} h_{22}+h_{13} f_{22} g_{23}-h_{13} f_{23} g_{22}\\
\\
M_{346} = f_{13} g_{22} h_{33}-f_{13} g_{33} h_{22}-g_{13} f_{22} h_{33}+g_{13} f_{33} h_{22}+h_{13} f_{22} g_{33}-h_{13} f_{33} g_{22}\\
\\
M_{356} = f_{13} g_{23} h_{33}-f_{13} g_{33} h_{23}-g_{13} f_{23} h_{33}+g_{13} f_{33} h_{23}+h_{13} f_{23} g_{33}-h_{13} f_{33} g_{23}\\
\\
M_{456} = f_{22} g_{23} h_{33}-f_{22} g_{33} h_{23}-g_{22} f_{23} h_{33}+g_{22} f_{33} h_{23}+h_{22} f_{23} g_{33}-h_{22} f_{33} g_{23}\\
\end{array}
\]
\smallskip\\
This time there is again a single Plucker relation

\begin{align}
\nonumber M_{1 2 3} M_{4 5 6} - M_{1 2 4} M_{3 5 6} + M_{1 2 5} M_{3 4 6} - M_{1 2 6} M_{3 4 5} + M_{1 3 4} M_{2 5 6} \\
\nonumber \\
- M_{1 3 5} M_{2 4 6} + M_{1 3 6} M_{2 4 5} + M_{1 4 5} M_{2 3 6} - M_{1 4 6} M_{2 3 5} + M_{1 5 6} M_{2 3 4} = 0
\end{align}
\smallskip\\
In this respect, the case $n = 3, r = 2$ is analogous to the case $n = 2, r = 4$. Be it coincidence or not, this is not the only parallel between these two cases: see the chapter 5 about integral discriminants. Resultant is a homogeneous polynomial of degree $4$ in minors:

\begin{center}
$R_{3|2}\{f,  g,  h\} = M_{1 2 3} M_{1 4 5} M_{1 5 6} M_{4 5 6}-M_{1 2 3} M_{1 4 5} M_{2 3 6} M_{4 5 6}-M_{1 2 3} M_{1 4 5} M_{3 4 5} M_{3 5 6}+M_{1 2 3} M_{1 4 5} M_{3 4 6}^2-M_{1 2 3} M_{1 4 6}^2 M_{4 5 6}+M_{1 2 3} M_{1 4 6} M_{2 3 5} M_{4 5 6}+M_{1 2 3} M_{1 4 6} M_{2 4 5} M_{3 5 6}-2 M_{1 2 3} M_{1 4 6} M_{2 4 6} M_{3 4 6}-M_{1 2 3} M_{1 5 6} M_{2 3 4} M_{4 5 6}+M_{1 2 3} M_{1 5 6} M_{2 4 6}^2+M_{1 2 3} M_{2 3 4} M_{2 4 6} M_{3 5 6}+M_{1 2 3} M_{2 3 4} M_{3 4 5} M_{3 5 6}-M_{1 2 3} M_{2 3 4} M_{3 4 6}^2-M_{1 2 3} M_{2 3 5} M_{2 4 5} M_{3 5 6}+M_{1 2 3} M_{2 3 5} M_{2 4 6} M_{3 4 6}+M_{1 2 3} M_{2 3 6} M_{2 4 5} M_{3 4 6}-M_{1 2 3} M_{2 3 6} M_{2 4 6}^2-M_{1 2 3} M_{2 3 6} M_{2 4 6} M_{3 4 5}-M_{1 2 4} M_{1 2 6} M_{1 5 6} M_{4 5 6}+M_{1 2 4} M_{1 2 6} M_{2 4 6} M_{3 5 6}+M_{1 2 4} M_{1 2 6} M_{3 4 5} M_{3 5 6}-M_{1 2 4} M_{1 2 6} M_{3 4 6}^2+M_{1 2 4} M_{1 3 5} M_{2 3 6} M_{4 5 6}-M_{1 2 4} M_{1 3 5} M_{2 4 6} M_{3 5 6}+M_{1 2 4} M_{1 3 6} M_{1 4 6} M_{4 5 6}-M_{1 2 4} M_{1 3 6} M_{2 3 5} M_{4 5 6}+M_{1 2 4} M_{1 3 6} M_{2 4 6} M_{3 4 6}-M_{1 2 4} M_{1 4 6}^2 M_{3 5 6}+M_{1 2 4} M_{1 4 6} M_{1 5 6} M_{3 4 6}+M_{1 2 4} M_{1 4 6} M_{2 3 5} M_{3 5 6}-M_{1 2 4} M_{1 4 6} M_{2 3 6} M_{3 4 6}-M_{1 2 4} M_{1 5 6}^2 M_{2 4 6}-M_{1 2 4} M_{1 5 6} M_{2 3 5} M_{3 4 6}+M_{1 2 4} M_{1 5 6} M_{2 3 6} M_{2 4 6}+M_{1 2 4} M_{1 5 6} M_{2 3 6} M_{3 4 5}+M_{1 2 5} M_{1 2 6} M_{1 4 6} M_{4 5 6}-M_{1 2 5} M_{1 2 6} M_{2 4 5} M_{3 5 6}+M_{1 2 5} M_{1 2 6} M_{2 4 6} M_{3 4 6}+M_{1 2 5} M_{1 3 4} M_{1 5 6} M_{4 5 6}-M_{1 2 5} M_{1 3 4} M_{2 3 6} M_{4 5 6}-M_{1 2 5} M_{1 3 4} M_{3 4 5} M_{3 5 6}+M_{1 2 5} M_{1 3 4} M_{3 4 6}^2-M_{1 2 5} M_{1 3 5} M_{1 4 6} M_{4 5 6}+M_{1 2 5} M_{1 3 5} M_{2 4 5} M_{3 5 6}-M_{1 2 5} M_{1 3 5} M_{2 4 6} M_{3 4 6}+M_{1 2 5} M_{1 3 6} M_{2 3 4} M_{4 5 6}-M_{1 2 5} M_{1 3 6} M_{2 4 5} M_{3 4 6}+M_{1 2 5} M_{1 3 6} M_{2 4 6} M_{3 4 5}+M_{1 2 5} M_{1 4 6}^2 M_{3 4 6}-M_{1 2 5} M_{1 4 6} M_{1 5 6} M_{2 4 6}-M_{1 2 5} M_{1 4 6} M_{1 5 6} M_{3 4 5}-M_{1 2 5} M_{1 4 6} M_{2 3 4} M_{3 5 6}+M_{1 2 5} M_{1 4 6} M_{2 3 6} M_{2 4 6}+M_{1 2 5} M_{1 5 6}^2 M_{2 4 5}+M_{1 2 5} M_{1 5 6} M_{2 3 4} M_{3 4 6}-M_{1 2 5} M_{1 5 6} M_{2 3 6} M_{2 4 5}-M_{1 2 6}^2 M_{1 4 5} M_{4 5 6}+M_{1 2 6}^2 M_{2 4 5} M_{3 4 6}-M_{1 2 6}^2 M_{2 4 6}^2-M_{1 2 6}^2 M_{2 4 6} M_{3 4 5}-M_{1 2 6} M_{1 3 4} M_{1 4 6} M_{4 5 6}+M_{1 2 6} M_{1 3 4} M_{2 3 5} M_{4 5 6}-M_{1 2 6} M_{1 3 4} M_{2 4 6} M_{3 4 6}+M_{1 2 6} M_{1 3 5} M_{1 4 5} M_{4 5 6}-M_{1 2 6} M_{1 3 5} M_{2 3 4} M_{4 5 6}+M_{1 2 6} M_{1 3 5} M_{2 4 6}^2+M_{1 2 6} M_{1 4 5}^2 M_{3 5 6}-3 M_{1 2 6} M_{1 4 5} M_{1 4 6} M_{3 4 6}+M_{1 2 6} M_{1 4 5} M_{1 5 6} M_{2 4 6}+M_{1 2 6} M_{1 4 5} M_{1 5 6} M_{3 4 5}+2 M_{1 2 6} M_{1 4 6}^2 M_{2 4 6}+M_{1 2 6} M_{1 4 6}^2 M_{3 4 5}-2 M_{1 2 6} M_{1 4 6} M_{1 5 6} M_{2 4 5}+M_{1 2 6} M_{1 4 6} M_{2 3 4} M_{3 4 6}-M_{1 2 6} M_{1 4 6} M_{2 3 5} M_{2 4 6}-M_{1 2 6} M_{1 5 6} M_{2 3 4} M_{2 4 6}-M_{1 2 6} M_{1 5 6} M_{2 3 4} M_{3 4 5}+M_{1 2 6} M_{1 5 6} M_{2 3 5} M_{2 4 5}-M_{1 3 4}^2 M_{1 5 6} M_{4 5 6}+M_{1 3 4}^2 M_{2 4 6} M_{3 5 6}+M_{1 3 4}^2 M_{3 4 5} M_{3 5 6}-M_{1 3 4}^2 M_{3 4 6}^2+M_{1 3 4} M_{1 3 5} M_{1 4 6} M_{4 5 6}-M_{1 3 4} M_{1 3 5} M_{2 4 5} M_{3 5 6}+M_{1 3 4} M_{1 3 5} M_{2 4 6} M_{3 4 6}-M_{1 3 4} M_{1 3 6} M_{1 4 5} M_{4 5 6}+M_{1 3 4} M_{1 3 6} M_{2 4 5} M_{3 4 6}-M_{1 3 4} M_{1 3 6} M_{2 4 6}^2-M_{1 3 4} M_{1 3 6} M_{2 4 6} M_{3 4 5}+2 M_{1 3 4} M_{1 4 5} M_{1 4 6} M_{3 5 6}-M_{1 3 4} M_{1 4 5} M_{1 5 6} M_{3 4 6}-M_{1 3 4} M_{1 4 5} M_{2 3 5} M_{3 5 6}+M_{1 3 4} M_{1 4 5} M_{2 3 6} M_{3 4 6}-2 M_{1 3 4} M_{1 4 6}^2 M_{3 4 6}+3 M_{1 3 4} M_{1 4 6} M_{1 5 6} M_{2 4 6}+M_{1 3 4} M_{1 4 6} M_{1 5 6} M_{3 4 5}+M_{1 3 4} M_{1 4 6} M_{2 3 5} M_{3 4 6}-M_{1 3 4} M_{1 4 6} M_{2 3 6} M_{2 4 6}-M_{1 3 4} M_{1 4 6} M_{2 3 6} M_{3 4 5}-M_{1 3 4} M_{1 5 6}^2 M_{2 4 5}-M_{1 3 5} M_{1 4 5}^2 M_{3 5 6}+M_{1 3 5} M_{1 4 5} M_{1 4 6} M_{3 4 6}+M_{1 3 5} M_{1 4 5} M_{2 3 4} M_{3 5 6}-M_{1 3 5} M_{1 4 5} M_{2 3 6} M_{2 4 6}-M_{1 3 5} M_{1 4 6}^2 M_{2 4 6}-M_{1 3 5} M_{1 4 6} M_{2 3 4} M_{3 4 6}+M_{1 3 5} M_{1 4 6} M_{2 3 6} M_{2 4 5}+M_{1 3 6} M_{1 4 5}^2 M_{3 4 6}-M_{1 3 6} M_{1 4 5} M_{1 4 6} M_{2 4 6}-M_{1 3 6} M_{1 4 5} M_{1 4 6} M_{3 4 5}-M_{1 3 6} M_{1 4 5} M_{2 3 4} M_{3 4 6}+M_{1 3 6} M_{1 4 5} M_{2 3 5} M_{2 4 6}+M_{1 3 6} M_{1 4 6}^2 M_{2 4 5}+M_{1 3 6} M_{1 4 6} M_{2 3 4} M_{2 4 6}+M_{1 3 6} M_{1 4 6} M_{2 3 4} M_{3 4 5}-M_{1 3 6} M_{1 4 6} M_{2 3 5} M_{2 4 5}-M_{1 4 5}^2 M_{1 5 6}^2+M_{1 4 5}^2 M_{1 5 6} M_{2 3 6}+2 M_{1 4 5} M_{1 4 6}^2 M_{1 5 6}-M_{1 4 5} M_{1 4 6}^2 M_{2 3 6}-M_{1 4 5} M_{1 4 6} M_{1 5 6} M_{2 3 5}+M_{1 4 5} M_{1 5 6}^2 M_{2 3 4}-M_{1 4 6}^4+M_{1 4 6}^3 M_{2 3 5}-M_{1 4 6}^2 M_{1 5 6} M_{2 3 4}+M_{2 5 6} M_{1 2 3} M_{1 4 6} M_{3 4 5}-M_{2 5 6} M_{1 2 3} M_{1 5 6} M_{2 4 5}-M_{2 5 6} M_{1 2 3} M_{2 3 4} M_{3 4 6}+M_{2 5 6} M_{1 2 3} M_{2 3 6} M_{2 4 5}-M_{2 5 6} M_{1 2 4} M_{1 2 6} M_{3 4 6}+M_{2 5 6} M_{1 2 4} M_{1 3 5} M_{3 4 6}-M_{2 5 6} M_{1 2 4} M_{1 3 6} M_{3 4 5}+M_{2 5 6} M_{1 2 4} M_{1 4 6} M_{1 5 6}-M_{2 5 6} M_{1 2 4} M_{1 4 6} M_{2 3 6}+M_{2 5 6} M_{1 2 6}^2 M_{2 4 5}+M_{2 5 6} M_{1 2 6} M_{1 3 4} M_{3 4 5}-M_{2 5 6} M_{1 2 6} M_{1 3 5} M_{2 4 5}-M_{2 5 6} M_{1 2 6} M_{1 4 5} M_{1 4 6}+M_{2 5 6} M_{1 2 6} M_{1 4 6} M_{2 3 4}-M_{2 5 6} M_{1 3 4}^2 M_{3 4 6}+M_{2 5 6} M_{1 3 4} M_{1 3 6} M_{2 4 5}-M_{2 5 6} M_{1 3 4} M_{1 4 5} M_{1 5 6}+M_{2 5 6} M_{1 3 4} M_{1 4 5} M_{2 3 6}-M_{2 5 6} M_{1 3 4} M_{1 4 6}^2+M_{2 5 6} M_{1 3 5} M_{1 4 5} M_{1 4 6}-M_{2 5 6} M_{1 3 6} M_{1 4 5} M_{2 3 4}$
\end{center}
This time it is not a determinant of some matrix (or, at least, such matrix is not yet found). Instead, it is a Pfaffian of the following $8 \times 8$ antisymmetric matrix, made of the Plucker coordinates:

\begin{align}
A = {\fontsize{3pt}{0pt}
\left( \begin{array}{ccccccccc}
0 & M_{3 5 6} & M_{4 5 6} & M_{2 4 6} & M_{1 5 6} & M_{1 4 6} & M_{2 5 6} & M_{3 4 6} \\
\\
-M_{3 5 6} & 0  & -M_{3 4 6} & M_{1 4 6} & M_{1 3 6} & M_{1 2 6} & M_{2 3 6} & M_{1 5 6}-M_{2 3 6} \\
\\
-M_{4 5 6} & M_{3 4 6} & 0 & M_{2 4 5} & M_{1 4 6} & M_{1 4 5} & M_{2 4 6} & M_{3 4 5} \\
\\
-M_{2 4 6} & -M_{1 4 6} & -M_{2 4 5} & 0 & M_{1 3 4} & M_{1 2 4} & M_{2 3 4}-M_{1 4 5} & -M_{2 3 4} \\
\\
-M_{1 5 6} & -M_{1 3 6} & -M_{1 4 6} & -M_{1 3 4} & 0 & M_{1 2 3} & -M_{1 2 6} & M_{1 2 6}-M_{1 3 5} \\
\\
-M_{1 4 6} & -M_{1 2 6} & -M_{1 4 5} & -M_{1 2 4} & -M_{1 2 3} & 0 & M_{1 3 4}-M_{1 2 5} & -M_{1 3 4} \\
\\
-M_{2 5 6} & -M_{2 3 6} & -M_{2 4 6} & -M_{2 3 4}+M_{1 4 5} & M_{1 2 6} & -M_{1 3 4}+M_{1 2 5} & 0 & M_{1 4 6}-M_{2 3 5} \\
\\
-M_{3 4 6} & -M_{1 5 6}+M_{2 3 6} & -M_{3 4 5} & M_{2 3 4} & -M_{1 2 6}+M_{1 3 5} & M_{1 3 4} & -M_{1 4 6}+M_{2 3 5} & 0
\end{array} \right)
}
\end{align}
\smallskip\\
The Pfaffian is equal to

\begin{align}
R_{3|2}\{f, g, h\} = {\rm pfaff} A = \epsilon_{i_1 \ldots i_8} A_{i_1 i_2} A_{i_3 i_4} A_{i_5 i_6} A_{i_7 i_8}
\label{Pfaff32}
\end{align}
\smallskip\\
where $\epsilon$ is the completely antisymmetric tensor (simply the antisymmetrisation over $i_1, \ldots, i_8$). The Pfaffian is also the same, as the square root of determinant, since for antisymmetric matrices determinant is always an exact square. Observation (\ref{Pfaff32}) was done in \cite{Eisenbud}, where the variables $M_{ijk}$ are being interpreted as coordinates on a Grassmanian manifold and then, using advanced homological algebra, this and similar formulas are being obtained. Generally speaking, resultant is expected to be a polynomial of degree $r^{n-1}$ in the Plucker coordinates $M_{i_1 \ldots i_n}$. It is an open problem to find an explicit formula for this polynomial.

\subsection{Formulas of Sylvester + Bezout type }

By combining the techniques of Sylvester and Bezout, it is possible to sufficiently enlarge the number of cases, where resultant is expressed as some polynomial (usually determinant) without any divisions. The basic example here is $n = 3, r = 2$. In this case, the system of quadratic equations has a form

\begin{align}
\left\{ \begin{array}{c}
f(x_1, x_2, x_3) = f_{11} x_1^2 + f_{12} x_1 x_2 + f_{13} x_1 x_3 + f_{22} x_2^2 + f_{23} x_2 x_3 + f_{33} x_3^2 = 0 \\
\noalign{\medskip}g(x_1, x_2, x_3) = g_{11} x_1^2 + g_{12} x_1 x_2 + g_{13} x_1 x_3 + g_{22} x_2^2 + g_{23} x_2 x_3 + g_{33} x_3^2 = 0 \\
\noalign{\medskip}h(x_1, x_2, x_3) = h_{11} x_1^2 + h_{12} x_1 x_2 + h_{13} x_1 x_3 + h_{22} x_2^2 + h_{23} x_2 x_3 + h_{33} x_3^2 = 0 \\
\end{array} \right. \label{theform32}
\end{align}
\smallskip\\
It can be represented as a $3 \times 6$ rectangular matrix. Obviously, another $3 \times 6$ matrix is necessary to complete a square $6 \times 6$ matrix. This complementary matrix is constructed quite elegantly, making use of Jacobian

$$J(x_1, x_2, x_3) = \det\left( \dfrac{\partial f_i}{\partial x_j} \right), \ \ \ f_1 \equiv f, f_2 \equiv g, f_3 \equiv h$$
\smallskip\\
which is a homogeneous polynomial of degree $3$ in $x_1, x_2, x_3$ and at the same time of degree $3$ in coefficients of $f, g$ and $h$. Jacobian can be expressed through Plucker variables $M_{ijk}$ as follows:

\begin{center}
$J(x_1  x_2  x_3) = M_{1 2 3} x_1^3+(-2 M_{1 3 4}+M_{1 2 5}) x_1^2 x_2+(2 M_{1 2 6}-M_{1 3 5}) x_1^2 x_3+(-M_{2 3 4}+2 M_{1 4 5}) x_1 x_2^2+(-M_{2 3 5}+4 M_{1 4 6}) x_1 x_2 x_3+(-M_{2 3 6}+2 M_{1 5 6}) x_1 x_3^2+M_{2 4 5} x_2^3+(2 M_{2 4 6}+M_{3 4 5}) x_2^2 x_3+(M_{2 5 6}+2 M_{3 4 6}) x_2 x_3^2+M_{3 5 6} x_3^3 $
\end{center}
The derivatives of $J$ are homogeneous quadratic polynomials in $x_1, x_2, x_3$, just like the equations $f,g$ and $h$:

\[
{\fontsize{3pt}{0pt}\left( \begin{array}{c}
f \\
\\
g \\
\\
h \\
\\
\dfrac{\partial J}{\partial x_1} \\
\\
\dfrac{\partial J}{\partial x_2} \\
\\
\dfrac{\partial J}{\partial x_3} \\
\end{array} \right) =
 \left( \begin{array}{cccccccc}
f_{11} & f_{12} & f_{13} & f_{22} & f_{23} & f_{33} \\
\\
g_{11} & g_{12} & g_{13} & g_{22} & g_{23} & g_{33} \\
\\
h_{11} & h_{12} & h_{13} & h_{22} & h_{23} & h_{33} \\
\\
3 M_{1 2 3} & -4 M_{1 3 4}+2 M_{1 2 5} & -2 M_{1 3 5}+4 M_{1 2 6} & -M_{2 3 4}+2 M_{1 4 5} & -M_{2 3 5}+4 M_{1 4 6} & -M_{2 3 6}+2 M_{1 5 6} \\
\\
-2 M_{1 3 4}+M_{1 2 5} & 4 M_{1 4 5}-2 M_{2 3 4} & -M_{2 3 5}+4 M_{1 4 6} & 3 M_{2 4 5} & 2 M_{3 4 5}+4 M_{2 4 6} & M_{2 5 6}+2 M_{3 4 6} \\
\\
2 M_{1 2 6}-M_{1 3 5} & -M_{2 3 5}+4 M_{1 4 6} & -2 M_{2 3 6}+4 M_{1 5 6} & M_{3 4 5}+2 M_{2 4 6} & 2 M_{2 5 6}+4 M_{3 4 6} & 3 M_{3 5 6} \\
\\
\end{array} \right) \left( \begin{array}{c}
x_1^2 \\
\\
x_1 x_2 \\
\\
x_1 x_3 \\
\\
x_2^2 \\
\\
x_2 x_3 \\
\\
x_3^2 \\
\end{array} \right)}
\]
\smallskip\\
The matrix in the right hand side is half Sylvester, half Bezout. Its determinant is equal to the resultant:

\begin{align}
{\fontsize{3pt}{0pt} R_{3|2}\{f,g,h\} = \left| \begin{array}{cccccccc}
f_{11} & f_{12} & f_{13} & f_{22} & f_{23} & f_{33} \\
\\
g_{11} & g_{12} & g_{13} & g_{22} & g_{23} & g_{33} \\
\\
h_{11} & h_{12} & h_{13} & h_{22} & h_{23} & h_{33} \\
\\
3 M_{1 2 3} & -4 M_{1 3 4}+2 M_{1 2 5} & -2 M_{1 3 5}+4 M_{1 2 6} & -M_{2 3 4}+2 M_{1 4 5} & -M_{2 3 5}+4 M_{1 4 6} & -M_{2 3 6}+2 M_{1 5 6} \\
\\
-2 M_{1 3 4}+M_{1 2 5} & 4 M_{1 4 5}-2 M_{2 3 4} & -M_{2 3 5}+4 M_{1 4 6} & 3 M_{2 4 5} & 2 M_{3 4 5}+4 M_{2 4 6} & M_{2 5 6}+2 M_{3 4 6} \\
\\
2 M_{1 2 6}-M_{1 3 5} & -M_{2 3 5}+4 M_{1 4 6} & -2 M_{2 3 6}+4 M_{1 5 6} & M_{3 4 5}+2 M_{2 4 6} & 2 M_{2 5 6}+4 M_{3 4 6} & 3 M_{3 5 6} \\
\\
\end{array} \right|}
\end{align}
\smallskip\\
In a compact notation this formula can be written as

\begin{align}
R_{3|2}\{f,g,h \} = \det_{6 \times 6}\left\{ f, g, h, \dfrac{\partial J}{\partial x_1}, \dfrac{\partial J}{\partial x_2}, \dfrac{\partial J}{\partial x_3} \right\}
\end{align}
\smallskip\\
where $\det\{ \ldots \}$ stands for the determinant of the matrix, obtained by expanding the polynomials inside brackets in the monomial basis. The other possibility to design a square matrix is to take 9 polynomials $x_i f_j$, which are homogeneous polynomials of degree $3$, and complement them with the Jacobian itself, not its derivatives. In this case one gets a $10 \times 10$ matrix in the monomial basis:

\begin{align}
R_{3|2}\{f,g,h \} = \det_{10 \times 10}\left\{ x_1 f, x_1 g, x_1 h, x_2 f, x_2 g, x_2 h, x_3 f, x_3 g, x_3 h, J \right\}
\end{align}
\smallskip\\
According to this method, it is necessary to make the matrix square by adding a number of certain polynomials to the Sylvester block. Of course, the most non-trivial part of the method is construction of these polynomials. For higher $r > 2$, these polynomials are constructed with the help of

$${\cal J}(x_1, \ldots, x_n, p_1, \ldots, p_n) = \det\left( \dfrac{\partial f_i}{\partial x_j} + \dfrac{1}{r-1} \dfrac{\partial^2 f_i}{\partial x_j \partial x_k} p_k + \dfrac{1}{(r-1)(r-2)} \dfrac{\partial^3 f_i}{\partial x_j \partial x_k \partial x_l} p_k p_l + \ldots \right)$$
\smallskip\\
Jacobian-like quantity ${\cal J}$ can be viewed as a series in the additional variables $p_1, \ldots, p_n$:

$${\cal J} (x_1, \ldots, x_n, p_1, \ldots, p_n) = J(x_1, \ldots, x_n) + J_i(x_1, \ldots, x_n) p_i + \dfrac{1}{2!} J_{ij}(x_1, \ldots, x_n) p_i p_j + \ldots $$
\smallskip\\
Note, that $J(x_1, \ldots, x_n)$ is nothing but the ordinary Jacobian and $J_i(x_1, \ldots, x_n)$ are its derivatives, while quantities $J_{ij}(x_1, \ldots, x_n)$ are already new and non-trivial (in particular, they are not equal to the second derivatives of $J$). In the next case $n = 3, r = 3$ the Sylvester-Bezout formulas take form

\begin{align}
R_{3|3}\{f_1,f_2,f_3 \} = \det_{21 \times 21}\left\{ x_i x_j f_k, J_l \right\}
\end{align}
and

\begin{align}
R_{3|3}\{f_1,f_2,f_3 \} = \det_{15 \times 15}\left\{ x_i f_j, J_{kl} \right\}
\end{align}
\smallskip\\
Just like in the previous case, there exist two different formulas. In the next case $n = 3, r = 4$ we have

\begin{align}
R_{3|4}\{f_1,f_2,f_3 \} = \det_{36 \times 36}\left\{ x_i x_j x_k f_l, J_{pq} \right\}
\end{align}
and

\begin{align}
R_{3|4}\{f_1,f_2,f_3 \} = \det_{28 \times 28}\left\{ x_i x_j f_k, J_{lpq} \right\}
\end{align}
\smallskip\\
Generally, for $n = 3$ and $r = $ any, we have two formulas of Sylvester-Bezout type:

\begin{equation}
\addtolength{\fboxsep}{5pt}
\boxed{
\begin{gathered}
R_{3|r}\{f_1,f_2,f_3 \} = \det_{r(2r+1) \times r(2r+1)}\left\{ x_{i_1} \ldots x_{i_{r-1}} f_{j}, J_{k_1 \ldots k_{r-2}} \right\}
\end{gathered}
}\label{3r-1}
\end{equation}
and

\begin{equation}
\addtolength{\fboxsep}{5pt}
\boxed{
\begin{gathered}
R_{3|r}\{f_1,f_2,f_3 \} = \det_{r(2r-1) \times r(2r-1)}\left\{ x_{i_1} \ldots x_{i_{r-2}} f_{j}, J_{k_1 \ldots k_{r-1}} \right\}
\end{gathered}
}\label{3r-2}
\end{equation}
\smallskip\\
As for higher dimensions $n > 3$, an explicit formula is known only for $n = 4, r = 2$ where it has a form

\begin{equation}
\addtolength{\fboxsep}{5pt}
\boxed{
\begin{gathered}
R_{4|2}\{f_1,f_2,f_3,f_4 \} = \det_{20 \times 20}\left\{ x_i f_j, J_k \right\}
\end{gathered}
}\label{4-2}
\end{equation}
\smallskip\\
For higher $n$ and $r$, such formulas are still an open direction of research, see \cite{Determ} for more details.

\subsection{Formulas of Schur type }

Resultants also admit a very different kind of formulas, related to certain analytic structure of the logarithm of resultant, described in \cite{Analytic}. By an analytic structure we mean that logarithm of resultant of a system

\[
\left\{
\begin{array}{ccc}
f_1(x_1, \ldots, x_n) = 0 \\
\\
f_2(x_1, \ldots, x_n) = 0 \\
\\
\ldots \\
\\
f_n(x_1, \ldots, x_n) = 0 \\
\end{array}
\right.
\]
\smallskip\\
depends on polynomials $f_1, \ldots, f_n$ in a very specific way, which is much simpler to understand than for the resultant itself. Two formulas are used to express this statement.

The first is the operator formula

\begin{equation}
\addtolength{\fboxsep}{5pt}
\boxed{
\begin{gathered}
 \log \ R_{r_1, \ldots, r_n}(I - f) = \left. \left[ \ \sum\limits_{k = 0}^{\infty} \dfrac{r_1}{(r_1 k)!} ({\hat f_1})^{k} \ \right] \ldots \left[ \ \sum\limits_{k = 0}^{\infty} \dfrac{r_n}{(r_n k)!} ({\hat f_n})^{k} \ \right] \cdot \log \ \det(I - A) \right|_{A = 0}
\end{gathered}
}\label{LogRezEx}
\end{equation}
\smallskip\\
where $A$ is an auxillary $n \times n$ matrix, and ${\hat f}_1, {\hat f}_2, \ldots, {\hat f}_n$ are differential operators

\begin{align}
\begin{array}{ccc}
{\hat f_1} =  f_1 \left( \dfrac{\partial}{\partial A_{11}}, \dfrac{\partial}{\partial A_{12}}, \ldots, \dfrac{\partial}{\partial A_{1n}} \right) \\
\noalign{\medskip} {\hat f_2} =  f_2 \left( \dfrac{\partial}{\partial A_{21}}, \dfrac{\partial}{\partial A_{22}}, \ldots, \dfrac{\partial}{\partial A_{2n}} \right) \\
\noalign{\medskip} \ldots \\
\noalign{\medskip} {\hat f_n} =  f_n \left( \dfrac{\partial}{\partial A_{n1}}, \dfrac{\partial}{\partial A_{n2}}, \ldots, \dfrac{\partial}{\partial A_{nn}} \right) \\
\end{array}
\end{align}
\smallskip\\
associated with polynomials $f_1, f_2, \ldots, f_n$. The second is the contour integral formula:

\begin{equation}
\addtolength{\fboxsep}{5pt}
\boxed{
\begin{gathered}
\log R \big\{ f_1, f_2, \ldots, f_n \big\} = \oint \ldots \oint \log f_1 \ d\log f_2 \ \wedge \ldots \wedge \ d\log f_n + r_1 \log R \big\{ f_2,\ldots,f_n \big\} \Big|_{x_1 = 0}
\end{gathered}
}\label{M0}
\end{equation}
\smallskip\\
where the contour integral is $(n-1)$-fold and functions $f_i$ are taken with arguments $ f_i(1,z_1,\ldots,z_{n-1}) $. The contour of integration is assumed to encircle all the common roots of the system

\begin{align}
 f_k(1,z_1,\ldots,z_{n-1}) = 0, \ \ k = 2, \ldots, n
\end{align}
\smallskip\\
see \cite{Analytic} for more details. An explicit formula for $\log R$ follows after several iterations of (\ref{M0}): for $n = 2$

\begin{align}
 \log R_{r_1 r_2} \left\{ f_1, f_2 \right\} = \oint \log f_1(1,z) d \log f_2(1,z) + r_1 \log f_2(0,1)
\end{align}
\smallskip\\
for $n = 3$

\begin{align*}
\log R_{r_1 r_2 r_3} \left\{ f_1, f_2, f_3 \right\} =
& \oint \oint \log f_1(1,z_1,z_2) d \log f_2(1,z_1,z_2) \wedge d \log f_3(1,z_1,z_2) + \emph{}
\\ & r_1 \oint \log f_2(0,1,z) d \log f_3(0,1,z) + r_1 r_2 \log f_3(0,0,1)
\end{align*}
\smallskip\\
and so on. In this way, logarithm of the resultant is expressed through the simple contour integrals. To obtain $R$ from $\log R$, we exponentiate the latter: for example, for $n = 2$

\begin{align}
 R_{r_1 r_2} \left\{ f_1, f_2 \right\} = \exp \left( \oint \log f_1(1,z) d \log f_2(1,z) + r_1 \log f_2(0,1) \right)
\label{M11}
\end{align}
\smallskip\\
Using this formula, it is straightforward to express the resultant of two polynomials through their roots: if

\begin{align}
 f_1(1,z) = a \prod\limits_{i = 1}^{r_1} (z - \alpha_i)
\end{align}

\begin{align}
 f_2(1,z) = b \prod\limits_{i = 1}^{r_2} (z - \beta_i) \end{align}
\smallskip\\
then

\begin{align}
R_{r_1 r_2} \left\{ f_1, f_2 \right\} = \exp \left( \sum\limits_{i = 1}^{r_1} \sum\limits_{j = 1}^{r_2} \oint \log (z - \alpha_i) d \log (z - \beta_j) + r_2 \log a + r_1 \log b \right) = a^{r_{2}} b^{r_{1}} \prod\limits_{i = 1}^{r_1} \prod\limits_{j = 1}^{r_2} ( \beta_{j} - \alpha_i )
\end{align}
\smallskip\\
Exponential formula (\ref{M11}) and its direct analogues for $n > 2$ have several advantages. The most important are simplicity and explicitness: such formulas have a transparent structure, which is easy to understand and memorize. Another one is universality: they can be written for any $n$, not just for low values. However, there is one serious drawback as well. Resultant is known to be a homogeneous polynomial of degree $$ d_i = \dfrac{r_1 r_2 \ldots r_n}{r_i} $$ in coefficients of the $i$-th equation $f_i$, and of total degree $$ d = d_1 + \ldots + d_n = r_1 r_2 \ldots r_n \left( \dfrac{1}{r_1} + \dfrac{1}{r_2} + \ldots + \dfrac{1}{r_n} \right) $$ in coefficients of all equations, but in exponential formula the polynomial nature of the resultant is not explicit. Therefore, it is not fully convenient for practical calculations of resultants (although it is conceptually adequate and can be used in theoretical considerations).

\pagebreak

To turn these formulas into a computational tool, one can apply (\ref{M0}) to a \emph{shifted} system

\begin{align}
 \left\{ \begin{array}{c}
{\tilde f}_1( {\vec x}) = (x_1)^{r_1} - \lambda_1 f_1( {\vec x}) \\
\noalign{\medskip}{\tilde f}_2( {\vec x}) = (x_2)^{r_2} - \lambda_2 f_2( {\vec x}) \\
\noalign{\medskip}\ldots\\
\noalign{\medskip}{\tilde f}_n( {\vec x}) = (x_n)^{r_n} - \lambda_n f_n( {\vec x}) \\
\end{array} \right.
\end{align}
\smallskip\\
and then expand the shifted logarithms in the integrands into Taylor series in powers of the "spectral parameters" $\lambda_i$. Then, one obtains the following series expansion

\begin{align}
\log R \big\{ {\tilde f}_1, \ldots, {\tilde f}_n \big\} = - \sum\limits_{k_1 = 0}^{\infty} \ldots \sum\limits_{k_n = 0}^{\infty} T_{k_1 k_2 \ldots k_n}\big\{ f_1, \ldots, f_n \big\} \cdot \lambda_1^{k_1} \lambda_2^{k_2} \ldots \lambda_n^{k_n}
\label{MTr}
\end{align}
\smallskip\\
(a minus sign is just a convention) where particular Taylor components $T_{k_1 k_2 \ldots k_n}$ are homogeneous polynomial expressions of degree $k_i$ in coefficients of $f_i$. We call them \emph{traces} of a non-linear system $f_1( {\vec x} ), \ldots, f_n( {\vec x} )$. By exponentiating, one obtains the shifted resultant:

\begin{align*}
R \big\{ {\tilde f}_1, \ldots, {\tilde f}_n \big\} = \exp \left( - \sum\limits_{k_1 = 0}^{\infty} \ldots \sum\limits_{k_n = 0}^{\infty} T_{k_1 k_2 \ldots k_n} \cdot \lambda_1^{k_1} \lambda_2^{k_2} \ldots \lambda_n^{k_n} \right)
\end{align*}
\smallskip\\
The original resultant $ R \big\{ f_1, \ldots, f_n \big\} $ is equal to $(-1)^{d}$ times the coefficient of $ \lambda_1^{d_1} \lambda_2^{d_2} \ldots \lambda_n^{d_n} $ in $R \big\{ {\tilde f}_1, \ldots, {\tilde f}_n \big\} $. We can extract it, expanding the right hand side in powers of $\lambda_i$:

\begin{align*}
\exp \left( - \sum\limits_{k_1 = 0}^{\infty} \ldots \sum\limits_{k_n = 0}^{\infty} T_{k_1 k_2 \ldots k_n} \cdot \lambda_1^{k_1} \lambda_2^{k_2} \ldots \lambda_n^{k_n} \right) = \sum\limits_{k_1 = 0}^{\infty} \ldots \sum\limits_{k_n = 0}^{\infty} {\cal P}_{k_1 k_2 \ldots k_n} \cdot \lambda_1^{k_1} \lambda_2^{k_2} \ldots \lambda_n^{k_n}
\end{align*}
\smallskip\\
Such power series expansion of $\exp \left( S(x) \right)$, where $S(x)$ is itself a power series (perhaps, of many variables), is often called a \emph{Schur expansion}. It implies the following relation between Taylor components:

\begin{align*}
{\cal P}_{k_1 k_2 \ldots k_n} \ = \ \sum\limits_{m = 1}^{k_1 + \ldots + k_n} \ \dfrac{(-1)^{m}}{m!} \ \sum\limits_{{\vec v}_1 + {\vec v}_2 + \ldots + {\vec v}_m = {\vec k}} \ T_{{\vec v}_1} \ T_{{\vec v}_2} \ \ldots \ T_{{\vec v}_m}
\end{align*}
\smallskip\\
where the sum is taken over all ordered partitions of a a vector ${\vec k} = (k_1, k_2, \ldots, k_n)$ into $m$ vectors, denoted as ${\vec v_1}, \ldots, {\vec v_m}$. An ordered partition is a way of writing a vector with integer components as a sum of vectors with integer components, where the order of the items is significant. Polynomials ${\cal P}_{k_1 k_2 \ldots k_n}(T)$ are often called multi-Schur polynomials. Several first multi-Schur polynomials are given by

\begin{align}
\begin{array}{cc}
{\cal P}_{1, 0}  = - T_{1, 0}\\
\\
{\cal P}_{2, 0}  = - T_{2, 0} + T_{1, 0}^2/2\\
\\
{\cal P}_{2, 1}  = - T_{2,1} + T_{2,0} T_{0,1} + T_{1,0} T_{1,1} - T_{1,0}^2 T_{0,1}/2\\
\\
{\cal P}_{2, 1, 0} = - T_{2,1,0} + T_{2,0,0} T_{0,1,0} + T_{1,0,0} T_{1,1,0} - T_{1,0,0}^2 T_{0,1,0}/2\\
\\
{\cal P}_{1, 1, 1} = - T_{1,1,1} + T_{1,0,0} T_{0,1,1} + T_{0,1,0} T_{1,0,1} + T_{1,1,0} T_{0,0,1} - T_{0,1,0} T_{0,0,1} T_{1,0,0}\\
\end{array}
\end{align}
\smallskip\\
Resultant is a certain multi-Schur polynomial of traces:

\begin{align}
R\big\{ f_1, \ldots, f_n \big\} = (-1)^d \ {\cal P}_{d_1,\ldots,d_n} = \sum\limits_{m = 1}^{d} \ \dfrac{(-1)^{m + d}}{m!} \ \sum\limits_{{\vec v}_1 + {\vec v}_2 + \ldots + {\vec v}_m = {\vec d}} \ T_{{\vec v}_1} \ T_{{\vec v}_2} \ \ldots \ T_{{\vec v}_m}
\label{M1}
\end{align}
\smallskip\\
The explicit formula for traces, obtained in \cite{Analytic}, has a form

\begin{align}
T_{\vec k} = T_{k_1 k_2 \ldots k_n} = \dfrac{1}{k_1 k_2 \ldots k_n} \cdot \sum\limits_{ r_{ij} } \ \det\limits_{2 \leq i,j \leq n} \big( \delta_{ij} r_{i} k_{i} - r_{ij} \big) \prod\limits_{i = 1}^{n} (f_i)^{k_i}_{r_{i1}, r_{i2}, \ldots ,r_{in}}
\label{M2}
\end{align}
\smallskip\\
where

\begin{align*}
 (f)^{k}_{j_1,j_2,\ldots, j_n} = \mbox{ coefficient of } x_1^{j_1} x_2^{j_2} \ldots x_n^{j_n} \mbox{ in } f(x_1,x_2,\ldots,x_n)^k
\end{align*}
\smallskip\\
and the sum goes over all non-negative integer $n \times n$ matrices $r_{ij}$ such that the sum of entries in any $i$-th row or $i$-th coloumn is fixed and equals $r_{i} k_{i}$. Formula (\ref{M2}) is valid for positive $k_1, \ldots, k_n$. The cases when some $k_i = 0$, also make no difficulty. If some $k_i = 0$, then

\begin{align}
T_{k_1,k_2,\ldots,k_n}\{ f_1,f_2,\ldots,f_n \} = r_i \left. T_{k_1,k_2,\ldots,k_{i-1},k_{i+1},\ldots,k_n}\{  f_1, f_2,\ldots, f_{i-1},f_{i+1},\ldots,f_{n}\} \right|_{x_i = 0}
\end{align}
\smallskip\\
where the trace in the right hand side is taken in variables $x_1, \ldots, x_{i-1}, x_{i+1}, \ldots, x_{n}$. It is an easy exercise to prove this statement, making a series expansion in (\ref{M0}). Similarly, operator approach gives \cite{Analytic}

\begin{align}
T_{k_1,k_2,\ldots,k_n}\{ f_1,f_2,\ldots,f_n \} = \left. \ \dfrac{r_1 ({\hat f_1})^{k_1}}{(r_1 k_1)!} \ldots \dfrac{r_n ({\hat f_n})^{k_n}}{(r_n k_n)!} \ \cdot \dfrac{ tr A^{r_1 k_1 + \ldots + r_n k_n} } {r_1 k_1 + \ldots + r_n k_n } \ \right|_{A = 0}
\label{M2alt}
\end{align}
\smallskip\\
Together, formulas \ref{M1} and \ref{M2} (or, alternatively, \ref{M2alt}) give an explicit algorithm to calculate resultants without any polynomial divisions. The absence of polynomial divisions should increase computational speed of the algorithm, if compared to algorithms with division such as Koszul complex.

To illustrate these general formulas, let us consider a particular calculation for $n = 3$ and $r = 2$, where the system of equations again has the form \ref{theform32}. The traces are calculated with (\ref{M2alt}), using differential operators

\[
\begin{array}{cc}
{\hat f} = f_{11} \left( \dfrac{ \partial }{\partial A_{11}} \right)^2 + f_{12} \dfrac{ \partial^2 }{\partial A_{11} \partial A_{12}} + f_{13} \dfrac{ \partial^2 }{\partial A_{11} \partial A_{13}} + f_{22} \left( \dfrac{ \partial }{\partial A_{12}} \right)^2 + f_{23} \dfrac{ \partial^2 }{\partial A_{12} \partial A_{13}} + f_{33} \left( \dfrac{ \partial }{\partial A_{13}} \right)^2  \\
\\
{\hat g} = g_{11} \left( \dfrac{ \partial }{\partial A_{21}} \right)^2 + g_{12} \dfrac{ \partial^2 }{\partial A_{21} \partial A_{22}} + g_{13} \dfrac{ \partial^2 }{\partial A_{21} \partial A_{23}} + g_{22} \left( \dfrac{ \partial }{\partial A_{22}} \right)^2 + g_{23} \dfrac{ \partial^2 }{\partial A_{22} \partial A_{23}} + g_{33} \left( \dfrac{ \partial }{\partial A_{23}} \right)^2 \\
\\
{\hat h} = h_{11} \left( \dfrac{ \partial }{\partial A_{31}} \right)^2 + h_{12} \dfrac{ \partial^2 }{\partial A_{31} \partial A_{32}} + h_{13} \dfrac{ \partial^2 }{\partial A_{31} \partial A_{33}} + h_{22} \left( \dfrac{ \partial }{\partial A_{32}} \right)^2 + h_{23} \dfrac{ \partial^2 }{\partial A_{32} \partial A_{33}} + h_{33} \left( \dfrac{ \partial }{\partial A_{33}} \right)^2
\end{array}
\]
\smallskip\\
and a few first traces are

\begin{align*}
& T_1 = 4 f_{1 1}+4 g_{2 2}+4 h_{3 3} \\
& \\
& T_2 = 4 f_{1 1}^2+4 g_{1 2} f_{1 2}+4 h_{1 3} f_{1 3}+8 g_{1 1} f_{2 2}+8 h_{1 1} f_{3 3}+4 g_{2 2}^2+8 g_{3 3} h_{2 2}+4 g_{2 3} h_{2 3}+4 h_{3 3}^2 \\
& \\
& T_3 = 4 f_{1 1}^3+6 g_{1 2} f_{1 1} f_{1 2}+6 h_{1 3} f_{1 1} f_{1 3}+12 g_{1 1} f_{1 1} f_{2 2}+12 h_{1 1} f_{1 1} f_{3 3}+6 g_{1 1} f_{1 2}^2+6 h_{1 1} f_{1 3}^2+3 g_{2 3} h_{1 3} f_{1 2}+ \emph{} \\ & 6 g_{3 3} h_{1 2} f_{1 2}+  6 g_{1 2} g_{2 2} f_{1 2}+3 g_{1 3} h_{2 3} f_{1 2}+3 g_{2 3} h_{1 2} f_{1 3}+6 h_{1 3} h_{3 3} f_{1 3}+6 g_{1 3} h_{2 2} f_{1 3}+3 g_{1 2} h_{2 3} f_{1 3}+ 6 g_{1 2}^2 f_{2 2}+ \emph{} \\ & 12 g_{1 1} g_{2 2} f_{2 2}+6 g_{1 3} h_{1 3} f_{2 2}+12 g_{3 3} h_{1 1} f_{2 2}+6 g_{2 3} h_{1 1} f_{2 3}+3 g_{1 2} h_{1 3} f_{2 3}+ 9 g_{1 3} h_{1 2} f_{2 3}+6 g_{1 1} h_{2 3} f_{2 3}+6 h_{1 3}^2 f_{3 3}+\emph{} \\ & 6 g_{1 2} h_{1 2} f_{3 3}+12 g_{1 1} h_{2 2} f_{3 3}+12 h_{1 1} h_{3 3} f_{3 3}+12 g_{2 2} g_{3 3} h_{2 2}+4 g_{2 2}^3+6 g_{2 3} h_{2 3} h_{3 3}+6 g_{2 2} g_{2 3} h_{2 3}+\emph{} \\ & 6 g_{2 3}^2 h_{2 2}+6 g_{3 3} h_{2 3}^2+12 g_{3 3} h_{2 2} h_{3 3} +4 h_{3 3}^3
\end{align*}
and so on. The resultant has degree 12, and is expressed through traces as

$$ R_{3|2}\big\{f,g,h\big\} = P_{12} \left\{ \dfrac{- T_{k}(f)}{k} \right\} = $$
\begin{center}
$
-\dfrac{1}{12} T_{12}+\dfrac{1}{72} T_{6}^2+\dfrac{1}{35} T_{5} T_{7}+\dfrac{1}{32} T_{4} T_{8}+\dfrac{1}{27} T_{3} T_{9}+\dfrac{1}{20} T_{2} T_{10}+\dfrac{1}{11} T_{1} T_{11}-\dfrac{1}{384} T_{4}^3-\dfrac{1}{60} T_{3} T_{4} T_{5}-\dfrac{1}{108} T_{3}^2 T_{6}-\dfrac{1}{100} T_{2} T_{5}^2-\dfrac{1}{48} T_{2} T_{4} T_{6}-\dfrac{1}{42} T_{2} T_{3} T_{7}-\dfrac{1}{64} T_{2}^2 T_{8}-\dfrac{1}{30} T_{1} T_{5} T_{6}-\dfrac{1}{28} T_{1} T_{4} T_{7}-\dfrac{1}{24} T_{1} T_{3} T_{8}-\dfrac{1}{18} T_{1} T_{2} T_{9}-\dfrac{1}{20} T_{1}^2 T_{10}+\dfrac{1}{1944} T_{3}^4+\dfrac{1}{144} T_{2} T_{3}^2 T_{4}+\dfrac{1}{256} T_{2}^2 T_{4}^2+\dfrac{1}{120} T_{2}^2 T_{3} T_{5}+\dfrac{1}{288} T_{2}^3 T_{6}+\dfrac{1}{96} T_{1} T_{3} T_{4}^2+\dfrac{1}{90} T_{1} T_{3}^2 T_{5}+\dfrac{1}{40} T_{1} T_{2} T_{4} T_{5}+\dfrac{1}{36} T_{1} T_{2} T_{3} T_{6}+\dfrac{1}{56} T_{1} T_{2}^2 T_{7}+\dfrac{1}{100} T_{1}^2 T_{5}^2+\dfrac{1}{48} T_{1}^2 T_{4} T_{6}+\dfrac{1}{42} T_{1}^2 T_{3} T_{7}+\dfrac{1}{32} T_{1}^2 T_{2} T_{8}+\dfrac{1}{54} T_{1}^3 T_{9}-\dfrac{1}{864} T_{2}^3 T_{3}^2-\dfrac{1}{1536} T_{2}^4 T_{4}-\dfrac{1}{324} T_{1} T_{2} T_{3}^3-\dfrac{1}{96} T_{1} T_{2}^2 T_{3} T_{4}-\dfrac{1}{240} T_{1} T_{2}^3 T_{5}-\dfrac{1}{144} T_{1}^2 T_{3}^2 T_{4}-\dfrac{1}{128} T_{1}^2 T_{2} T_{4}^2-\dfrac{1}{60} T_{1}^2 T_{2} T_{3} T_{5}-\dfrac{1}{96} T_{1}^2 T_{2}^2 T_{6}-\dfrac{1}{120} T_{1}^3 T_{4} T_{5}-\dfrac{1}{108} T_{1}^3 T_{3} T_{6}-\dfrac{1}{84} T_{1}^3 T_{2} T_{7}-\dfrac{1}{192} T_{1}^4 T_{8}+\dfrac{1}{46080} T_{2}^6+\dfrac{1}{1152} T_{1} T_{2}^4 T_{3}+\dfrac{1}{288} T_{1}^2 T_{2}^2 T_{3}^2+\dfrac{1}{384} T_{1}^2 T_{2}^3 T_{4}+\dfrac{1}{972} T_{1}^3 T_{3}^3+\dfrac{1}{144} T_{1}^3 T_{2} T_{3} T_{4}+\dfrac{1}{240} T_{1}^3 T_{2}^2 T_{5}+\dfrac{1}{768} T_{1}^4 T_{4}^2+\dfrac{1}{360} T_{1}^4 T_{3} T_{5}+\dfrac{1}{288} T_{1}^4 T_{2} T_{6}+\dfrac{1}{840} T_{1}^5 T_{7}-\dfrac{1}{7680} T_{1}^2 T_{2}^5-\dfrac{1}{864} T_{1}^3 T_{2}^3 T_{3}-\dfrac{1}{864} T_{1}^4 T_{2} T_{3}^2-\dfrac{1}{768} T_{1}^4 T_{2}^2 T_{4}-\dfrac{1}{1440} T_{1}^5 T_{3} T_{4}-\dfrac{1}{1200} T_{1}^5 T_{2} T_{5}-\dfrac{1}{4320} T_{1}^6 T_{6}+\dfrac{1}{9216} T_{1}^4 T_{2}^4+\dfrac{1}{2880} T_{1}^5 T_{2}^2 T_{3}+\dfrac{1}{12960} T_{1}^6 T_{3}^2+\dfrac{1}{5760} T_{1}^6 T_{2} T_{4}+\dfrac{1}{25200} T_{1}^7 T_{5}-\dfrac{1}{34560} T_{1}^6 T_{2}^3-\dfrac{1}{30240} T_{1}^7 T_{2} T_{3}-\dfrac{1}{161280} T_{1}^8 T_{4}+\dfrac{1}{322560} T_{1}^8 T_{2}^2+\dfrac{1}{1088640} T_{1}^9 T_{3}-\dfrac{1}{7257600} T_{1}^{10} T_{2}+\dfrac{1}{479001600} T_{1}^{12}
$\smallskip\\
\end{center}
This is a rather compact formula for $R_{3|2}$. By substituting the expressions for traces, we obtain a polynomial expression in $f$, $g$ and $h$, which contains 21894 monomials and takes several dozens of A4 pages.

\section{Calculation of discriminant }

Discriminant is a particular case of resultant, associated with gradient systems of equations. Since discriminant is a function of a single homogeneous polynomial (instead of $n$ homogeneous polynomials in the resultant case) it is slightly simpler than resultant, has bigger symmetry and some specific methods of calculation. These discriminant-specific methods are the topic of this section.

\begin{figure}[t]
\begin{center}
\includegraphics[totalheight=130pt]{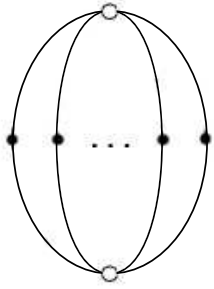}
\caption{Determinant of matrix $S_{ij}$, represented as a diagram of tensor contraction. Black 2-valent vertices represent tensor $S$, white n-valent vertices represent tensor $\epsilon$.}
\end{center}
\end{figure}

\subsection{Discriminant through invariants }

Let $S(x_1, \ldots, x_n)$ be a homogeneous polynomial of degree $r$ in $n$ variables. Discriminant of $S$ is a polynomial $SL(n)$ invariant of $S$. As such, it should be expressible as a polynomial in some simpler basis invariants, given by elegant diagram technique suggested in \cite{Nolinal}. The simplest example is of course the determinant, given by a single diagram at Fig.1. The simplest non-trivial (i.e. non-Gaussian) example is a $3$-form in $2$ variables:

$$ S(x_1,x_2) = S_{111} x_1^3 + 3 S_{112} x_1^2 x_2 + 3 S_{122} x_1 x_2^2 + S_{222} x_2^2 $$
\smallskip\\
By dimension counting, there is only one elementary invariant $I_{4}$ in this case, given by a diagram at Fig. 2.
\begin{figure}[t]
\begin{center}
\includegraphics[totalheight=160pt,clip]{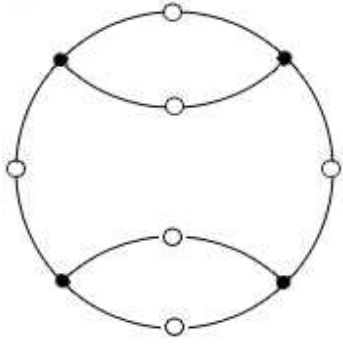}
\caption{The degree 4 invariant $I_4$ of a 3-form in 2 variables, represented as a diagram of tensor contraction. Black 3-valent vertices represent tensor $S$, white 2-valent vertices represent tensor $\epsilon$.}
\end{center}
\end{figure}
The subscript "4" stands for the degree of this invariant. In this paper we find it convenient to denote the elementary invariants of degree $k$ as $I_{k}$. It is straightforward to write the algebraic expression for the diagram:

$$ I_{4} = S_{i_1 i_2 i_3} S_{j_1 j_2 j_3} S_{k_1 k_2 k_3} S_{l_1 l_2 l_3} \epsilon^{i_1 j_1} \epsilon^{i_2 j_2} \epsilon^{k_1 l_1} \epsilon^{k_2 l_2} \epsilon^{i_3 k_3} \epsilon^{j_3 l_3} $$
\smallskip\\
Evaluating this sum, one gets the following explicit formula for $I_4$

$$ I_{4} = 2 S_{111}^2 S_{222}^2-12 S_{111} S_{112} S_{122} S_{222}+8 S_{111} S_{122}^3+8 S_{112}^3 S_{222}-6 S_{112}^2 S_{122}^2$$
\smallskip\\
which is nothing but the algebraic discriminant $ D_{2|3} $ of $S$:

\begin{equation}
\addtolength{\fboxsep}{5pt}
\boxed{
\begin{gathered}
D_{2|3} = I_4
\end{gathered}
}\label{D23}
\end{equation}
\smallskip\\
The next-to-simplest example is a $4$-form in $2$ variables, which can be written as

\begin{align*}
S(x_1,x_2) = S_{1111} x_1^4 + 4 S_{1112} x_1^3 x_2 + 6 S_{1122} x_1^2 x_2^2 + 4 S_{1222} x_1 x_2^3 + S_{2222} x_2^4
\end{align*}
\smallskip\\
By dimension counting, there are two elementary invariants in this case. They have relatively low degrees $2$ and $3$, denoted as $I_{2}$ and $I_3$ and given by diagrams at Fig. 3 and Fig. 4, respectively. Looking at the diagrams, it is straightforward to write algebraic expressions for $I_2, I_3$:
\begin{figure}[h]
\begin{center}
\includegraphics[totalheight=125pt,clip]{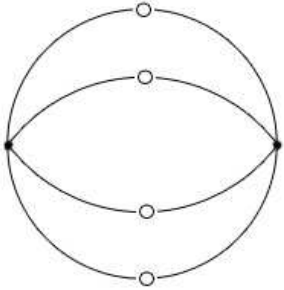}
\caption{The degree 2 invariant $I_2$ of a 4-form in 2 variables, represented as a diagram of tensor contraction. Black 4-valent vertices represent tensor $S$, white 2-valent vertices represent tensor $\epsilon$.}
\end{center}
\end{figure}

\begin{figure}[h]
\begin{center}
\includegraphics[totalheight=125pt,clip]{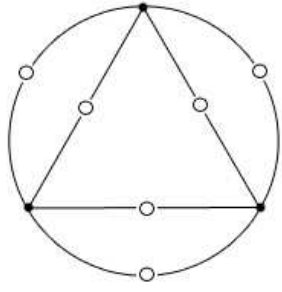}
\caption{The degree 3 invariant $I_3$ of a 4-form in 2 variables, represented as a diagram of tensor contraction. Black 4-valent vertices represent tensor $S$, white 2-valent vertices represent tensor $\epsilon$.}
\end{center}
\end{figure}

$$ I_{2} = S_{i_1 i_2 i_3 i_4} S_{j_1 j_2 j_3 j_4} \epsilon^{i_1 j_1} \epsilon^{i_2 j_2} \epsilon^{i_3 j_3} \epsilon^{i_4 j_4}$$

$$ I_{3} = S_{i_1 i_2 i_3 i_4} S_{j_1 j_2 j_3 j_4} S_{k_1 k_2 k_3 k_4} \epsilon^{i_1 j_1} \epsilon^{i_2 j_2} \epsilon^{i_3 k_1} \epsilon^{i_4 k_2} \epsilon^{j_3 k_3} \epsilon^{j_4 k_4}$$
\smallskip\\
Evaluating these sums, one gets the following explicit formulas for $I_2, I_3$:

$$ I_{2} = 2 S_{1111} S_{2222} - 8 S_{1112} S_{1222} + 6 S_{1122}^2$$

$$ I_{3} = 6 S_{1111} S_{1122} S_{2222}-6 S_{1111} S_{1222}^2-6 S_{1112}^2 S_{2222}+12 S_{1112} S_{1122} S_{1222}-6 S_{1122}^3$$
\smallskip\\
The algebraic discriminant $D_{2|4}$, just like any other $SL(2)$-invariant function of $S$, is a function of $I_2,I_3$:

\begin{align*}
\boxed{ D_{2|4} = I_2^3 - 6 I_{3}^2 } = 8 S_{1111}^3 S_{2222}^3-96 S_{1111}^2 S_{1112} S_{1222} S_{2222}^2-144 S_{1111}^2 S_{1122}^2 S_{2222}^2+ 432 S_{1111}^2 S_{1122} S_{1222}^2 S_{2222}
\end{align*}
\vspace{-1ex}
\begin{align}
\nonumber & -216 S_{1111}^2 S_{1222}^4+432 S_{1111} S_{1112}^2 S_{1122} S_{2222}^2-48 S_{1111} S_{1112}^2 S_{1222}^2 S_{2222}-1440 S_{1111} S_{1112} S_{1122}^2 S_{1222} S_{2222} \\ & \nonumber + 648 S_{1111} S_{1122}^4 S_{2222} + 864 S_{1111} S_{1112} S_{1122} S_{1222}^3 +864 S_{1112}^3 S_{1122} S_{1222} S_{2222} +288 S_{1112}^2 S_{1122}^2 S_{1222}^2 \\ & -432 S_{1111} S_{1122}^3 S_{1222}^2-216 S_{1112}^4 S_{2222}^2-512 S_{1112}^3 S_{1222}^3-432 S_{1112}^2 S_{1122}^3 S_{2222}
\label{D24}
\end{align}
Our third example is a $5$-form in $2$ variables, which can be written as

$$ S(x_1,x_2) = S_{11111} x_1^5 + 5 S_{11112} x_1^4 x_2 + 10 S_{11122} x_1^3 x_2^2 + 10 S_{11222} x_1^2 x_2^3 + 5 S_{12222} x_1 x_2^4 + S_{22222} x_2^5 $$
By dimension counting, there are three elementary invariants in this case. They have degrees $4$, $8$ and $12$, denoted as $I_{4}$, $I_{8}$ and $I_{12}$ and given by diagrams at Fig. 5, Fig. 6 and Fig. 7.
\begin{figure}[h]
\begin{center}
\includegraphics[totalheight=130pt,clip]{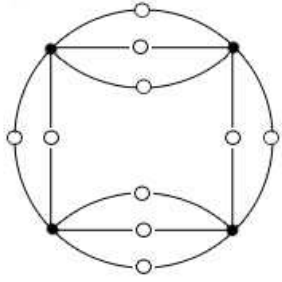}
\caption{The degree 4 invariant $I_4$ of a 5-form in 2 variables, represented as a diagram of tensor contraction. Black 5-valent vertices represent tensor $S$, white 2-valent vertices represent tensor $\epsilon$.}
\end{center}
\end{figure}

\begin{figure}[t]
\begin{center}
\includegraphics[totalheight=120pt,clip]{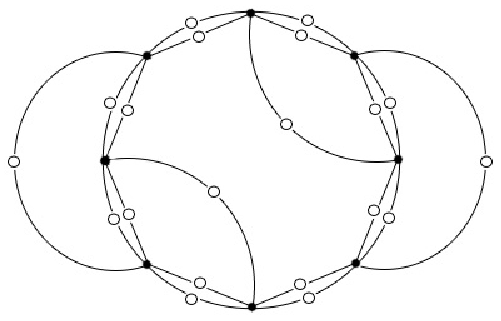}
\caption{The degree 8 invariant $I_8$ of a 5-form in 2 variables, represented as a diagram of tensor contraction. Black 5-valent vertices represent tensor $S$, white 2-valent vertices represent tensor $\epsilon$.}
\end{center}
\end{figure}

\begin{figure}[t]
\begin{center}
\includegraphics[totalheight=230pt,clip]{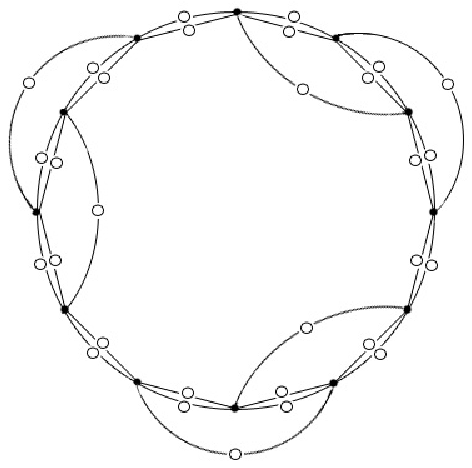}
\caption{The degree 12 invariant $I_{12}$ of a 5-form in 2 variables, represented as a diagram of tensor contraction. Black 5-valent vertices represent tensor $S$, white 2-valent vertices represent tensor $\epsilon$.}
\end{center}
\end{figure}
Looking at the diagrams, it is straightforward to write an expression for $I_{4}$

$$ I_{4} = S_{i_1 i_2 i_3 i_4 i_5} S_{j_1 j_2 j_3 j_4 j_5} S_{k_1 k_2 k_3 k_4 k_5} S_{l_1 l_2 l_3 l_4 l_5} \epsilon^{i_1 j_1} \epsilon^{i_2 j_2} \epsilon^{i_3 j_3} \epsilon^{i_4 k_4} \epsilon^{i_5 k_5} \epsilon^{j_4 l_4} \epsilon^{j_5 l_5} \epsilon^{k_1 l_1} \epsilon^{k_2 l_2} \epsilon^{k_3 l_3} $$
\smallskip\\
and equally straightforward to write expressions for $I_{8}, I_{12}$. Evaluating the contraction, one gets a formula

\begin{align*}
I_{4} \ = \ & 2 S_{11111}^2 S_{22222}^2-20 S_{11111} S_{11112} S_{12222} S_{22222}+8 S_{11111} S_{11122} S_{11222} S_{22222}+ \emph{} \\ & 32 S_{11111} S_{11122} S_{12222}^2-24 S_{11111} S_{11222}^2 S_{12222}+32 S_{11112}^2 S_{11222} S_{22222}+18 S_{11112}^2 S_{12222}^2-\emph{} \\ &24 S_{11112} S_{11122}^2 S_{22222}-152 S_{11112} S_{11122} S_{11222} S_{12222}+96 S_{11112} S_{11222}^3+96 S_{11122}^3 S_{12222}-64 S_{11122}^2 S_{11222}^2
\end{align*}
\smallskip\\
and similar formulas for $I_{8}, I_{12}$ -- they are quite lengthy and we do not present them here. The algebraic discriminant $D_{2|5}$, just like any other $SL(2)$-invariant function of $S$, is a function of $I_4,I_8,I_{12}$:

\begin{equation}
\addtolength{\fboxsep}{5pt}
\boxed{
\begin{gathered}
D_{2|5} = I_4^2 - 64 I_{8}
\end{gathered}
}\label{D25}
\end{equation}
\smallskip\\
Our final example is a $3$-form in $3$ variables, which can be written as

\begin{center}
$ S(x_1,x_2,x_3) = S_{111} x_1^3 + 3 S_{112} x_1^2 x_2 + 3 S_{113} x_1^2 x_3 + S_{222} x_2^3 + 3 S_{122} x_1x_2^2 + 3 S_{223} x_2^2x_3 + S_{333} x_3^3 + 3 S_{133} x_1x_3^2 + 3 S_{233} x_2x_3^2 + 6 S_{123} x_1x_2x_3  $
\end{center}
By dimension counting, there are two elementary invariants in this case. They have degrees $4$ and $6$, denoted as $I_{4}, I_{6}$ and given by diagrams at Fig. 8, Fig. 9.

\begin{figure}[t]
\begin{center}
\includegraphics[totalheight=200pt,clip]{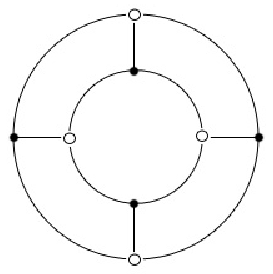}
\caption{The degree 4 invariant $I_{4}$ of a 3-form in 3 variables, represented as a diagram of tensor contraction. Black 3-valent vertices represent tensor $S$, white 3-valent vertices represent tensor $\epsilon$. }
\end{center}
\end{figure}

\begin{figure}[t]
\begin{center}
\includegraphics[totalheight=230pt,clip]{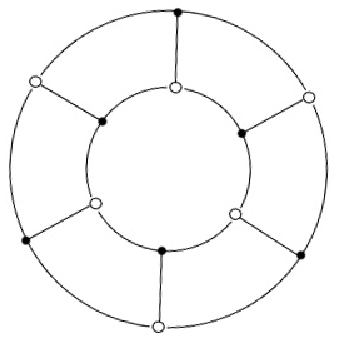}
\caption{The degree 6 invariant $I_{6}$ of a 3-form in 3 variables, represented as a diagram of tensor contraction. Black 3-valent vertices represent tensor $S$, white 3-valent vertices represent tensor $\epsilon$. }
\end{center}
\end{figure}
Looking at the diagrams, it is straightforward to write the algebraic expressions for $I_{4}$ and $I_{6}$:

$$
I_4 = S_{i_1 i_2 i_3} S_{j_1 j_2 j_3} S_{k_1 k_2 k_3} S_{l_1 l_2 l_3} \epsilon^{i_1 j_1 k_1} \epsilon^{i_2 j_2 l_2} \epsilon^{i_3 k_3 l_3} \epsilon^{l_1 k_2 j_3}
$$

$$
I_6 = S_{i_1 i_2 i_3} S_{j_1 j_2 j_3} S_{k_1 k_2 k_3} S_{l_1 l_2 l_3} S_{m_1 m_2 m_3} S_{s_1 s_2 s_3} \epsilon^{i_1 k_1 l_1}\epsilon^{i_2 j_2 s_2}\epsilon^{j_1 k_2 m_1}\epsilon^{l_2 m_2 k_3}\epsilon^{m_3 s_3 j_3}\epsilon^{l_3 i_3 s_1}
$$
\smallskip\\
Evaluating these sums, one gets the following explicit formulas for $I_4, I_6$:

\begin{center}
$ I_{4} = 6 S_{123}^4-12 S_{122} S_{123}^2 S_{133}+6 S_{122}^2 S_{133}^2+6 S_{113} S_{123} S_{133} S_{222}-12 S_{113} S_{123}^2 S_{223}-6 S_{113} S_{122} S_{133} S_{223}+18 S_{113} S_{122} S_{123} S_{233}-6 S_{113} S_{122}^2 S_{333}+6 S_{113}^2 S_{223}^2-6 S_{113}^2 S_{222} S_{233}-6 S_{112} S_{133}^2 S_{222}+18 S_{112} S_{123} S_{133} S_{223}-12 S_{112} S_{123}^2 S_{233}-6 S_{112} S_{122} S_{133} S_{233}+6 S_{112} S_{122} S_{123} S_{333}-6 S_{112} S_{113} S_{223} S_{233}+6 S_{112} S_{113} S_{222} S_{333}+6 S_{112}^2 S_{233}^2-6 S_{112}^2 S_{223} S_{333}-6 S_{111} S_{133} S_{223}^2+6 S_{111} S_{133} S_{222} S_{233}+6 S_{111} S_{123} S_{223} S_{233}-6 S_{111} S_{123} S_{222} S_{333}-6 S_{111} S_{122} S_{233}^2+6 S_{111} S_{122} S_{223} S_{333}$
\end{center}

\begin{center}
$ I_6 = 48 S_{123}^6-144 S_{122} S_{123}^4 S_{133}+144 S_{122}^2 S_{123}^2 S_{133}^2-48 S_{122}^3 S_{133}^3+72 S_{113} S_{123}^3 S_{133} S_{222}-144 S_{113} S_{123}^4 S_{223}-72 S_{113} S_{122} S_{123} S_{133}^2 S_{222}+72 S_{113} S_{122} S_{123}^2 S_{133} S_{223}+216 S_{113} S_{122} S_{123}^3 S_{233}+72 S_{113} S_{122}^2 S_{133}^2 S_{223}-216 S_{113} S_{122}^2 S_{123} S_{133} S_{233}-72 S_{113} S_{122}^2 S_{123}^2 S_{333}+72 S_{113} S_{122}^3 S_{133} S_{333}+18 S_{113}^2 S_{133}^2 S_{222}^2-72 S_{113}^2 S_{123} S_{133} S_{222} S_{223}+144 S_{113}^2 S_{123}^2 S_{223}^2-72 S_{113}^2 S_{123}^2 S_{222} S_{233}+72 S_{113}^2 S_{122} S_{133} S_{223}^2-36 S_{113}^2 S_{122} S_{133} S_{222} S_{233}-216 S_{113}^2 S_{122} S_{123} S_{223} S_{233}+144 S_{113}^2 S_{122} S_{123} S_{222} S_{333}+162 S_{113}^2 S_{122}^2 S_{233}^2-144 S_{113}^2 S_{122}^2 S_{223} S_{333}-48 S_{113}^3 S_{223}^3+72 S_{113}^3 S_{222} S_{223} S_{233}-24 S_{113}^3 S_{222}^2 S_{333}-72 S_{112} S_{123}^2 S_{133}^2 S_{222}+216 S_{112} S_{123}^3 S_{133} S_{223}-144 S_{112} S_{123}^4 S_{233}+72 S_{112} S_{122} S_{133}^3 S_{222}-216 S_{112} S_{122} S_{123} S_{133}^2 S_{223}+72 S_{112} S_{122} S_{123}^2 S_{133} S_{233}+72 S_{112} S_{122} S_{123}^3 S_{333}+72 S_{112} S_{122}^2 S_{133}^2 S_{233}-72 S_{112} S_{122}^2 S_{123} S_{133} S_{333}-36 S_{112} S_{113} S_{133}^2 S_{222} S_{223}-216 S_{112} S_{113} S_{123} S_{133} S_{223}^2+360 S_{112} S_{113} S_{123} S_{133} S_{222} S_{233}+72 S_{112} S_{113} S_{123}^2 S_{223} S_{233}-216 S_{112} S_{113} S_{123}^2 S_{222} S_{333}+36 S_{112} S_{113} S_{122} S_{133} S_{223} S_{233}-108 S_{112} S_{113} S_{122} S_{133} S_{222} S_{333}-216 S_{112} S_{113} S_{122} S_{123} S_{233}^2+360 S_{112} S_{113} S_{122} S_{123} S_{223} S_{333}-36 S_{112} S_{113} S_{122}^2 S_{233} S_{333}+72 S_{112} S_{113}^2 S_{223}^2 S_{233}-144 S_{112} S_{113}^2 S_{222} S_{233}^2+72 S_{112} S_{113}^2 S_{222} S_{223} S_{333}+162 S_{112}^2 S_{133}^2 S_{223}^2-144 S_{112}^2 S_{133}^2 S_{222} S_{233}-216 S_{112}^2 S_{123} S_{133} S_{223} S_{233}+144 S_{112}^2 S_{123} S_{133} S_{222} S_{333}+144 S_{112}^2 S_{123}^2 S_{233}^2-72 S_{112}^2 S_{123}^2 S_{223} S_{333}+72 S_{112}^2 S_{122} S_{133} S_{233}^2-36 S_{112}^2 S_{122} S_{133} S_{223} S_{333}-72 S_{112}^2 S_{122} S_{123} S_{233} S_{333}+18 S_{112}^2 S_{122}^2 S_{333}^2+72 S_{112}^2 S_{113} S_{223} S_{233}^2-144 S_{112}^2 S_{113} S_{223}^2 S_{333}+72 S_{112}^2 S_{113} S_{222} S_{233} S_{333}-48 S_{112}^3 S_{233}^3+72 S_{112}^3 S_{223} S_{233} S_{333}-24 S_{112}^3 S_{222} S_{333}^2-24 S_{111} S_{133}^3 S_{222}^2+144 S_{111} S_{123} S_{133}^2 S_{222} S_{223}-72 S_{111} S_{123}^2 S_{133} S_{223}^2-216 S_{111} S_{123}^2 S_{133} S_{222} S_{233}+72 S_{111} S_{123}^3 S_{223} S_{233}+120 S_{111} S_{123}^3 S_{222} S_{333}-144 S_{111} S_{122} S_{133}^2 S_{223}^2+72 S_{111} S_{122} S_{133}^2 S_{222} S_{233}+360 S_{111} S_{122} S_{123} S_{133} S_{223} S_{233}-72 S_{111} S_{122} S_{123} S_{133} S_{222} S_{333}-72 S_{111} S_{122} S_{123}^2 S_{233}^2-216 S_{111} S_{122} S_{123}^2 S_{223} S_{333}-144 S_{111} S_{122}^2 S_{133} S_{233}^2+72 S_{111} S_{122}^2 S_{133} S_{223} S_{333}+144 S_{111} S_{122}^2 S_{123} S_{233} S_{333}-24 S_{111} S_{122}^3 S_{333}^2+72 S_{111} S_{113} S_{133} S_{223}^3-108 S_{111} S_{113} S_{133} S_{222} S_{223} S_{233}+36 S_{111} S_{113} S_{133} S_{222}^2 S_{333}-72 S_{111} S_{113} S_{123} S_{223}^2 S_{233}+144 S_{111} S_{113} S_{123} S_{222} S_{233}^2-72 S_{111} S_{113} S_{123} S_{222} S_{223} S_{333}-36 S_{111} S_{113} S_{122} S_{223} S_{233}^2+72 S_{111} S_{113} S_{122} S_{223}^2 S_{333}-36 S_{111} S_{113} S_{122} S_{222} S_{233} S_{333}-36 S_{111} S_{112} S_{133} S_{223}^2 S_{233}+72 S_{111} S_{112} S_{133} S_{222} S_{233}^2-36 S_{111} S_{112} S_{133} S_{222} S_{223} S_{333}-72 S_{111} S_{112} S_{123} S_{223} S_{233}^2+144 S_{111} S_{112} S_{123} S_{223}^2 S_{333}-72 S_{111} S_{112} S_{123} S_{222} S_{233} S_{333}+72 S_{111} S_{112} S_{122} S_{233}^3-108 S_{111} S_{112} S_{122} S_{223} S_{233} S_{333}+36 S_{111} S_{112} S_{122} S_{222} S_{333}^2+18 S_{111}^2 S_{223}^2 S_{233}^2-24 S_{111}^2 S_{223}^3 S_{333}-24 S_{111}^2 S_{222} S_{233}^3+36 S_{111}^2 S_{222} S_{223} S_{233} S_{333}-6 S_{111}^2 S_{222}^2 S_{333}^2$ \end{center} \vspace{-1ex}
The algebraic discriminant $D_{3|3}$, just like any other $SL(3)$-invariant function of $S$, is a function of $I_4$ and $I_6$:

\begin{equation}
\addtolength{\fboxsep}{5pt}
\boxed{
\begin{gathered}
D_{3|3} = 32 I_4^3 + 3 I_6^2
\end{gathered}
}\label{D33}
\end{equation}
\smallskip\\
When expanded, discriminant $D_{3|3}$ contains 2040 monomials. See the appendix of the book version of \cite{Nolinal}, where it is written explicitly. Formula (\ref{D33}) is a remarkably concise expression of this disriminant through a pair of invariants, given by beautiful diagrams Fig.13 and Fig.14. It is an open problem to generalize this type of formulas -- eqs. (\ref{D23}), (\ref{D24}), (\ref{D25}) and (\ref{D33}) -- to higher $n$ and $r$. For more information on invariant theory, see \cite{InvTheor}.

\subsection{Discriminant through Newton polytopes }

A homogeneous polynomial of degree $r$ in $n$ variables

$$ S(x_1,x_2,\ldots,x_n) = \sum\limits_{a_1 + a_2 + \ldots + a_n = r } s_{a_1,a_2 \ldots a_n } x_{1}^{a_1} x_{2}^{a_2} \ldots x_{n}^{a_n} $$
\smallskip\\
can be visualized as a collection of points with coordinates $(a_1, \ldots, a_n)$ in the $n$-dimensional Euclidean space. The convex hull of these points is a $n$-dimensional convex polytope, called Newton polytope of $S$. It is known, that different triangulations and cellular decompositions of this polytope are in correspondence with different monomials in the discriminant of $S$. We do not go into details on this topic, see \cite{GKZ} for more details.

\subsection{The symmetric case }

An important special case is the case of symmetric polynomials $S(x_1, \ldots, x_n)$, i.e, those which are invariant under permutations of variables $x_1, \ldots, x_n$. It turns out that exactly for such polynomials discriminant greatly simplifies and factorizes into many smaller parts. A few examples for low degrees of $S$ can be given to illustrate this phenomenon. The simplest example is the case of degree two, because it is completely treatable by methods of linear algebra. A homogeneous symmetric polynomial of degree two should have a form

\begin{align}
S(x_1, \ldots, x_n) = C_{2} p_2 + C_{11} p_1^2
\end{align}
\smallskip\\
where $C_{2}, C_{11}$ are two arbitrary parameters and $p_k = \sum_i x_i^k$. To calculate the discriminant, we take derivatives:

\begin{align}
\dfrac{\partial S}{\partial x_i} = 2 C_{2} x_i + 2 C_{11} p_1 = 0
\label{Sys2}
\end{align}
\smallskip\\
The simplest option is just to write a matrix and calculate its determinant by usual rules:

\begin{align}
{\cal D}_{n|2}\big(C_{2}, C_{11}\big) = \det\limits_{n \times n} \left(
\begin{array}{cccc}
C_{2} + C_{11} & C_{11} & \ldots & C_{11} \\
\\
C_{11} & C_{2} + C_{11} & \ldots & C_{11} \\
\\
\ldots \\
C_{11} & C_{11} & \ldots & C_{2} + C_{11} \\
\end{array} \right) = C_{2}^{n-1} \big( C_2 + n C_{11} \big)
\end{align}
\smallskip\\
As one can see, the discriminant is highly reducible: factorises into simple elementary constitutients. In fact, this is a general property of symmetric polynomials. The next-to-simplest case is a homogeneous symmetric polynomial of degree $3$ in $n$ variables, which has a form

\begin{align}
S(x_1, \ldots, x_n) = C_{3} p_3 + C_{21} p_2 p_1 + C_{111} p_1^3
\end{align}
\smallskip\\
and contains three parameters $C_{3}, C_{21}, C_{111}$. As shown in \cite{Symmetrics}, the discriminant of this polynomial equals

{\fontsize{9pt}{0pt}
\begin{align}
{\cal D}_{n|3}\big(C_{3}, C_{21}, C_{111}\big) = \big( B_3 \big)^{(n-3)2^{n-1}} \prod\limits_{k = 0}^{n - 1} \left( \left( \dfrac{n-2k}{9n} \right)^2 B_1 B_3^2 + \dfrac{4k(n-k)}{27 n^2} B_2^3 \right)^{\dfrac{(n-1)!}{k!(n-1-k)!}}
\end{align}
\smallskip\\
where

\begin{align}
\left\{
\begin{array}{lll}
B_{1} = n^{2}C_{111} + n C_{21} + C_{3},\\
\\
B_{2} = n C_{21} + 3 C_{3},\\
\\
B_{3} = C_{3},
\end{array}
\right.
\end{align}
\smallskip\\
This formula is remarkably concise and, most importantly, it contains $n$ just as a parameter and is valid in any dimension $n$. This property makes it possible to study the large $n$ asymptotics and various continous limits. Moreover, the above examples for $r = 2$ and $r = 3$ can be generalized to arbitrary degree. Generally, a symmetric polynomial homogeneous of degree $r$ in $n$ variables has a form

\begin{align}
S(x_1, \ldots, x_n) = \sum\limits_{|Y| = r} C_Y p_Y
\label{GenSymm}
\end{align}
\smallskip\\
where the sum goes over Young diagrams (partitions) $Y: k_1 \geq k_2 \ldots \geq 0$ of fixed sum $|Y| = k_1 + k_2 + \ldots = r$, and $p_Y = \prod_{i} p_{Y_i}$. The number of free parameters is equal to the number of partitions of degree $r$, which we denote as $P(r)$. For example, $P(2) = 2$ corresponds to partitions $(2)$ and $(1,1)$. Similarly $P(3) = 3$ corresponds to partitions $(3), (2,1)$ and $(1,1,1)$. Several first values of $P(r)$ are

\begin{align}
\begin{array}{c|cccccccccccccc}
r & 1 & 2 & 3 & 4 & 5 & 6 & 7 & 8 & 9 & 10 \\
\\
P(r) & 1 & 2 & 3 & 5 & 7 & 11 & 15 & 22 & 30 & 42 \\
\end{array}
\end{align}
\smallskip\\
The generating function for $P(r)$ is given by

\begin{align}
\sum\limits_{r = 0}^{\infty} P(r) q^r = \prod\limits_{n = 1}^{\infty} \dfrac{1}{1 - q^n}
\end{align}
\smallskip\\
A symmetric homogeneous polynomial of degree $r$ has $P(r)$ independent coefficients $C_Y$. Its discriminant is a function of these coefficients, which we denote ${\cal D}_{n|r} \left( C \right)$. According to \cite{Symmetrics}, this discriminant is equal to

\begin{equation}
\addtolength{\fboxsep}{5pt}
\boxed{
\begin{gathered}
{\cal D}_{n|r} \left( C \right) = {\rm const} \cdot C_r^{\beta_{n|r}} \ \prod\limits_{M_1 + \ldots + M_{r-1} = n} \Big( d_{M}\left( C \right) \Big)^{ \dfrac{\#_M!}{(r-1)!} \dfrac{(M_1 + \ldots + M_{r-1})!}{M_1! \ldots M_{r-1}!}  }
\end{gathered}
}\label{DiscrimDecomp}
\end{equation}
\smallskip\\
where the product is taken over all decompositions of $n$ into $r - 1$ nonnegative parts and $\#_M$ is the number of zeroes among $M_1, \ldots, M_{r-1}$. The degree $\beta_{n|r}$ is fixed by the total degree of the discriminant

\begin{align}
\beta_{n|r} = \deg {\cal D}_{n|r} - \sum\limits_{M_1 + \ldots + M_{r-1} = n} \dfrac{\#_M! \deg d_{M}}{(r-1)!} \dfrac{(M_1 + \ldots + M_{r-1})!}{M_1! \ldots M_{r-1}!}
\label{BetaDegree}
\end{align}
and

\begin{align}
d_{M}\left( C \right) = R \left\{ \ \sum\limits_{i} P^{(i)}_M, \ \ \sum\limits_{i < j} P^{(i j)}_M, \ \ \sum\limits_{i < j < k} P^{(i j k)}_M, \ \ \ldots \ \right\}
\end{align}
\smallskip\\
with

\begin{align}
P^{(ij)}_{M} = \dfrac{ \det\limits_{2 \times 2}\left( \begin{array}{cc} 1 & P^{(i)}_M \\ \\ 1 & P^{(j)}_M \end{array} \right) }{ \det\limits_{2 \times 2} \left( \begin{array}{cc} 1 & y_i \\ \\ 1 & y_j \end{array} \right) } \ , \ \ \ \ \ \ \ P^{(ijk)}_{M} = \dfrac{ \det\limits_{3 \times 3} \left( \begin{array}{ccc} 1 & y_i & P^{(i)}_M \\ \\ 1 & y_j & P^{(j)}_M \\ \\ 1 & y_k & P^{(k)}_M \end{array} \right) }{ \det\limits_{3 \times 3} \left( \begin{array}{ccc} 1 & y_i & y_i^2 \\ \\ 1 & y_j & y_j^2 \\ \\ 1 & y_k & y_k^2 \end{array} \right) } \ , \ \ \  \ \ \ \ \ldots
\label{2and3levelquantities}
\end{align}
\smallskip\\
and so on, generally

\begin{align}
P^{(i_1 \ldots i_k)}_{M} = \dfrac{ \det\limits_{k \times k} \left( \begin{array}{ccccc} 1 & y_{i_1} & \ldots & y_{i_1}^{k-2} & P^{(i_1)}_M \\ \\ 1 & y_{i_2} & \ldots & y_{i_2}^{k-2} & P^{(i_2)}_M \\ \\ \ldots \\ \\ 1 & y_{i_k} & \ldots & y_{i_k}^{k-2} & P^{(i_k)}_M \\ \\\end{array} \right) }{ \det\limits_{k \times k} \left( \begin{array}{ccccc} 1 & y_{i_1} & \ldots & y_{i_1}^{k-2} & y_{i_1}^{k-1} \\ \\ 1 & y_{i_2} & \ldots & y_{i_2}^{k-2} & y_{i_2}^{k-1} \\ \\ \ldots \\ \\ 1 & y_{i_k} & \ldots & y_{i_k}^{k-2} & y_{i_k}^{k-1} \\ \\\end{array} \right) }
\end{align}
\smallskip\\
Note, that each of quantities $P^{(i_1 \ldots i_k)}_{M}$ is a polynomial of degree $r - k$ in variables $y_i$. They are defined only for pairwise distinct upper indices $i_1, \ldots, i_k$ and are symmetric in them. For $k \geq r$ they are undefined: one can not choose $r$ distinct items out of $(r-1)$. As one can see, each particular factor $d_{M}\left( S \right)$ can be computed as a resultant in no more than in $(r - 1)$ variables. This allows to calculate discriminants of symmetric polynomials on the practical level even for $n >> r$.

\section{Integral discriminants and applications}

\subsection{Non-Gaussian integrals and Ward identities}

Resultants and discriminants from the previous sections are purely algebraic (or combinatorial) objects -- i.e, they are polynomials, possibly lengthy and complicated, but still finite linear combinations of monomials. The next level of complexity in non-linear algebra is occupied by \emph{partition functions}, which are usually not polynomials and not even elementary functions. Still, they posess a number of remarkable properties and are related closely to resultants and discriminants.
The simplest partition functions of non-linear algebra are integral discriminants

\begin{align}
J_{n|r}\big(S\big) = \int dx_1 \ldots dx_n \ e^{-S(x_1, \ldots, x_n)}
\label{Intd}
\end{align}
\smallskip\\
For quadratic forms, diagonalisation and other linear-algebra methods can be used to obtain a simple answer $ J_{n|2}\big(S\big) = \big( \det S \big)^{-1/2}$, but for higher degrees $r$ such methods do not work. Instead, to calculate integral discriminants one needs to use a powerful technique of Ward identities, which are linear differential equations

\begin{align}
\left(\dfrac{\partial}{\partial s_{a_1, \ldots, a_n}} \dfrac{\partial}{\partial s_{b_1, \ldots, b_n}} -  \dfrac{\partial}{\partial s_{c_1, \ldots, c_n}} \dfrac{\partial}{\partial s_{d_1, \ldots, d_n}} \right) J_{n|r}(S) = 0, \ \ \ \ a_i + b_i = c_i + d_i
\label{Ward}
\end{align}
\smallskip\\
satisfied by integral discriminants. Validity of these identities follows simply from the fact, that differential operator in the left hand side of (\ref{Ward}) annihilates the integrand in (\ref{Intd}). Making use of the Ward identities, in \cite{IntDisc} integral discriminants were found in the simplest low-dimensional cases:

{\fontsize{9pt}{0pt}\[
\begin{array}{c|c|c|c|l}
n & r & \mbox{ Invariants } & \mbox{ Discriminant } D_{n|r} & \mbox{ First branch of the integral discriminant } J_{n|r} \\
& & & \\
\hline
& & & \\
2 & 2 & I_2 & I_2 & \ I_{2}^{-1/2} \\
& & & \\
& & & \\
2 & 3 & I_4 & I_2 &\ I_{4}^{-1/6} \\
& & & \\
& & & \\
2 & 4 & I_2, I_3 & I_2^3 - 6 I_3^2 &\ I_{2}^{-1/4} \cdot \sum\limits_{i = 0}^{\infty} \dfrac{1}{i!} \cdot \dfrac{(1/12)_{i} (5/12)_{i} }{(1/2)_{i}} \cdot \left( \dfrac{6I_{3}^2}{I_{2}^3} \right)^i \\
& & & \\
& & & \\
2 & 5 & I_4, I_8, I_{12} & I_4^2 - 64 I_8 & \ I_{4}^{-1/10} \cdot \sum\limits_{i,j = 0}^{\infty} \dfrac{1}{i!j!} \cdot \dfrac{(3/10)_{i + j} (1/10)_{2i + 3j} (1/10)_{j} }{(2/5)_{i + 2j} (3/5)_{i + 2j}} \cdot \left( \dfrac{16I_{8}}{I_{4}^2} \right)^i \left( \dfrac{128I_{12}}{3I_{4}^3} \right)^j  \\
& & & \\
& & & \\
3 & 2 & I_3 & I_3 & \ I_{3}^{-1/2} \\
& & & \\
3 & 3 & I_4, I_6 & 32 I_4^3 + 3 I_6^2 & \ I_{4}^{-1/4} \cdot \sum\limits_{i = 0}^{\infty} \dfrac{1}{i!} \cdot \dfrac{(1/12)_{i} (5/12)_{i} }{(1/2)_{i}} \cdot \left( -\dfrac{3I_{6}^2}{32 I_{4}^3} \right)^i \\
& & & \\
& & & \\
\ldots & \ldots & \ldots & \ldots & \ldots \\
\end{array}
\]}
where

$$
(a)_k \equiv \frac{\Gamma(a+k)}{\Gamma(a)} = a(a+1) \dotsb (a+k-1)
$$
\smallskip\\
Already by definition the integral (\ref{Intd}) is $SL(n)$-invariant and, consequently, depends only on the elementary $SL(n)$-invariants $I_k$, which were described in section 4.1. above. It is necessary to note, that $J_{n|r}$ is a multi-valued function: for one and the same $n|r$ there can be several branches, which satisfy one and the same equations (\ref{Ward}), but have different asymptotics. For example, for $n|r = 2|4$ integral discriminant has exactly two branches:

\begin{align}
J^{(1)}_{2|4}\big(S\big) = I_{2}^{-1/4} \ {}_{2}F_{1} \left( \left. \begin{array}{ccc} 1/12, 5/12 \\ 1/2 \end{array} \right| \ \dfrac{6I_{3}^2}{I_{2}^3} \ \right)  = I_{2}^{-1/4} \cdot \sum\limits_{i = 0}^{\infty} \dfrac{1}{i!} \cdot \dfrac{(1/12)_{i} (5/12)_{i} }{(1/2)_{i}} \cdot \left( \dfrac{6I_{3}^2}{I_{2}^3} \right)^i
\end{align}
\smallskip\\

\begin{align}
J^{(2)}_{2|4}\big(S\big) = I_{3} I_{2}^{-7/4} \ {}_{2}F_{1} \left( \left. \begin{array}{ccc} 7/12, 11/12 \\ 3/2 \end{array} \right| \ \dfrac{6I_{3}^2}{I_{2}^3} \ \right)  = I_{3} I_{2}^{-7/4} \cdot \sum\limits_{i = 0}^{\infty} \dfrac{1}{i!} \cdot \dfrac{(7/12)_{i} (11/12)_{i} }{(3/2)_{i}} \cdot \left( \dfrac{6I_{3}^2}{I_{2}^3} \right)^i
\end{align}
\smallskip\\
For the sake of brevity, only one of branches is presented in the table above -- the one with normalisation factor depending only on the invariant of lowest degree $I_{\rm min}$. Other branches are omitted. As one can see, non-Gaussian integrals posess a nice structure: they are all hypergeometric, i.e, coefficients in their series expansions are ratios of $\Gamma$-functions. As shown in \cite{IntDisc}, in all these cases integral discriminant $J_{n|r}(S)$ has singularities in the zero locus of the ordinary algebraic discriminant $D_{n|r}(S)$. This interesting relation between purely algebraic objects (discriminants) and non-Gaussian integrals deserves to be further investigated.

\subsection{Action-independence}

Actually, an even stronger statement is valid: not only

\begin{align}
\left(\dfrac{\partial}{\partial s_{a_1, \ldots, a_n}} \dfrac{\partial}{\partial s_{b_1, \ldots, b_n}} -  \dfrac{\partial}{\partial s_{c_1, \ldots, c_n}} \dfrac{\partial}{\partial s_{d_1, \ldots, d_n}} \right) \int dx_1 \ldots dx_n \ e^{-S(x_1, \ldots, x_n)} = 0, \ \ \ \ a_i + b_i = c_i + d_i
\end{align}
\smallskip\\
but also

\begin{align}
\left(\dfrac{\partial}{\partial s_{a_1, \ldots, a_n}} \dfrac{\partial}{\partial s_{b_1, \ldots, b_n}} -  \dfrac{\partial}{\partial s_{c_1, \ldots, c_n}} \dfrac{\partial}{\partial s_{d_1, \ldots, d_n}} \right) \int dx_1 \ldots dx_n \ f\Big( S(x_1, \ldots, x_n) \Big) = 0, \ \ \ \ a_i + b_i = c_i + d_i
\end{align}
\smallskip\\
with (almost) arbitrary choice of function $f(S)$. Again, the validity of these differential equations obviously follows from the fact, that differential operator in the left hand side annihilates the integrand. As one can see, two different integrals -- with integrands $e^{-S}$ and $f(S)$ -- satisfy one and the same Ward identities. By itself, this fact does not guarantee equality of these two integrals. However, both integrals are $SL(n)$ invariant and have one and the same degree of homogeneity: for any $f(S)$ we have

\begin{align}
\int dx_1 \ldots dx_n \ f\Big( \lambda S(x_1, \ldots, x_n) \Big) = \lambda^{-n/r} \int dx_1 \ldots dx_n \ f\Big( S(x_1, \ldots, x_n) \Big)
\end{align}
\smallskip\\
Together, these three properties -- common Ward identities, common degree of homogeneity and $SL(n)$ invariance -- can be considered as a proof (or, at least, as a strong evidence in favor) of equivalence of these two differently looking integrals: for arbitrary good function $f(S)$, we have

\begin{equation}
\addtolength{\fboxsep}{5pt}
\boxed{
\begin{gathered}
\int f \Big( S(x_1, \ldots, x_n) \Big) \ dx_1 \ldots dx_n = {\rm const}(f) \ \int e^{- S(x_1, \ldots, x_n)} \ dx_1 \ldots dx_n
\end{gathered}
}\label{Prop2}
\end{equation}
\smallskip\\
The constant here depends on the choice of $f$, but does not depend on $S$. Property (\ref{Prop2}) is sometimes called "action-independence" of integral discriminants. Even before specifying the class of good functions $f(S)$ and proving the action independence, let us consider a simple illustration with $f(S) = e^{-S^2}$:

$$
\int e^{- \big(S_{ij} x_i x_j\big)} \ d^n x = \int e^{- \big(\lambda_1 x_1^2 + \ldots + \lambda_n x_n^2\big)} \ d^n x = \dfrac{1}{\sqrt{\lambda_1 \ldots \lambda_n}} \int e^{- \big(x_1^2 + \ldots + x_n^2\big)} \ d^n x = \dfrac{{\rm const}}{\sqrt{\det S}}
$$

$$
\int e^{- \big(S_{ij} x_i x_j\big)^2} \ d^n x = \int e^{- \big(\lambda_1 x_1^2 + \ldots + \lambda_n x_n^2\big)^2} \ d^n x = \dfrac{1}{\sqrt{\lambda_1 \ldots \lambda_n}} \int e^{- \big(x_1^2 + \ldots + x_n^2\big)^2} \ d^n x = \dfrac{{\rm const^{\prime}}}{\sqrt{\det S}}
$$
\smallskip\\
As one can see, these two differently-looking integrals are indeed essentially one and the same (they are proportional with $S$-independent constant of proportionality). To specify the class of good functions and to prove (\ref{Prop2}), let us make a change of integration variables

$$x_1 = \rho, \ \ x_2 = \rho z_2, \ \ x_3 = \rho z_3, \ \ \ldots, x_n = \rho z_n $$
\smallskip\\
i.e. pass from homogeneous coordinates $x_i$ to non-homogeneous coordinates $z_i = x_i / x_1$. Then

$$ \int e^{- S(x_1, x_2, \ldots, x_n)} \ dx_1 \ldots dx_n = \int \rho^{n - 1} d\rho \int dz_2 \ldots dz_n \ e^{- \rho^r S(1, z_2, \ldots, z_n)} = $$

$$ = \left( \int \rho^{n - 1} e^{- \rho^r} d\rho \right) \cdot \int \dfrac{dz_2 \ldots dz_n}{S(1, z_2, \ldots, z_n)^{n/r}} $$
\smallskip\\
For the right hand side of (\ref{Prop2}) we have

$$ \int f\Big( S(x_1, x_2, \ldots, x_n) \Big) \ dx_1 \ldots dx_n = \int \rho^{n - 1} d\rho \int dz_2 \ldots dz_n \ f\Big( \rho^r S(1, z_2, \ldots, z_n) \Big) = $$

$$ = \left( \int \rho^{n - 1} f\Big( \rho^r \Big) d\rho \right) \cdot \int \dfrac{dz_2 \ldots dz_n}{S(1, z_2, \ldots, z_n)^{n/r}} $$
\smallskip\\
Both integrals over $\rho$ are just $S$-independent constants. As one can see, (\ref{Prop2}) is valid, iff these integrals

$$ \int \rho^{n - 1} e^{- \rho^r} d\rho \ \ \ \ {\rm and} \ \ \ \ \int \rho^{n - 1} f\Big( \rho^r \Big) d\rho $$
\smallskip\\
are finite over one and the same contour. This condition specifies the class of good functions $f(S)$. For example, all functions $ f(S) = \exp \big( - S^k \big) $ for $k > 0$ fall into this class.

\subsection{Evaluation of areas}

Action independence of integral discriminants allows to make different choices for $f(S)$. This explains why integral discriminants appear in a variety of differently looking problems. For example, we can take

\begin{align}
f(S) = \theta(1 - S) = \left\{ \begin{array}{ccc} 1, \ \ \ S \leq 1 \\ \\ 0, \ \ \ \ S > 1 \end{array} \right.
\end{align}
\smallskip\\
which makes sence only in the real setting, i.e, when the contour of integration is real (and the form $S$ is positive definite). This choice of $f(S)$ provides an important geometric interpretation of integral discriminants:

\begin{equation}
\addtolength{\fboxsep}{5pt}
\boxed{
\begin{gathered}
\int e^{- S(x_1, x_2, \ldots, x_n)} \ dx_1 \ldots dx_n \mathop{=}^{(60)-(61)} \int\limits_{S < 1} dx_1 \ldots dx_n = \mbox{ volume, bounded by the hypersurface S = 1}
\end{gathered}
}\label{BoundedVolume}
\end{equation}
\smallskip\\
This relation to bounded volume is a nice practical application of integral discriminant theory. As an example, let us consider the case of $J_{2|4}$. Take a particular algebraic curve

\begin{align}
S(x,y) = (x^2 + y^2)^2 + \epsilon x^2 y^2 = 1
\end{align}
\smallskip\\
The area, bounded by this curve, by definition can be calculated in polar coordinates as

\begin{align*}
V(\epsilon) = \int\limits_{0}^{2\pi} \dfrac{d \phi}{2 \sqrt{1 + \epsilon \cos^2 \phi \sin^2 \phi }} = \dfrac{4}{\sqrt{\epsilon + 4}} K\left( \dfrac{\epsilon}{\epsilon + 4} \right)
\end{align*}
\smallskip\\
where $K$ is the complete elliptic integral of the first kind \cite{Gradstein} and can be represented as

\begin{align}
V(\epsilon) = \dfrac{4}{\sqrt{\epsilon + 4}} K\left( \dfrac{\epsilon}{\epsilon + 4} \right) = \dfrac{2 \pi}{\sqrt{\epsilon + 4}} \ {}_{2}F_{1} \left( \left. \begin{array}{ccc} 1/2, 1/2 \\ 1 \end{array} \right| \  \dfrac{\epsilon}{\epsilon + 4} \ \right)
\end{align}
\smallskip\\
From the integral discriminant side, the homogeneous quartic form $S(x,y)$ has two invariants

\begin{align}
I_2 = \dfrac{8}{3} + \dfrac{2}{3} \epsilon + \dfrac{1}{6} \epsilon^2, \ \ \ \ \ I_3 = \dfrac{16}{9} + \dfrac{2}{3} \epsilon - \dfrac{1}{6} \epsilon^2 - \dfrac{1}{36} \epsilon^3
\end{align}
\smallskip\\
and the two branches of the integral discriminant are

\begin{align}
J^{(1)}_{2|4}\big(S\big) = \left( \dfrac{8}{3} + \dfrac{2}{3} \epsilon + \dfrac{1}{6} \epsilon^2 \right)^{-1/4} \ {}_{2}F_{1} \left( \left. \begin{array}{ccc} 1/12, 5/12 \\ 1/2 \end{array} \right| \ \dfrac{ (\epsilon + 2)^2 (\epsilon+8)^2 (\epsilon-4)^2}{(\epsilon^2 + 4 \epsilon + 16)^3} \ \right)
\end{align}

\begin{align}
J^{(2)}_{2|4}\big(S\big) = \left( \dfrac{16}{9} + \dfrac{2}{3} \epsilon - \dfrac{1}{6} \epsilon^2 - \dfrac{1}{36} \epsilon^3 \right) \left( \dfrac{8}{3} + \dfrac{2}{3} \epsilon + \dfrac{1}{6} \epsilon^2 \right)^{-7/4} \ {}_{2}F_{1} \left( \left. \begin{array}{ccc} 7/12, 11/12 \\ 3/2 \end{array} \right| \ \dfrac{ (\epsilon + 2)^2 (\epsilon+8)^2 (\epsilon-4)^2}{(\epsilon^2 + 4 \epsilon + 16)^3} \ \right)
\end{align}
\smallskip\\
As noticed in \cite{IntDisc}, only one linear combination of these two branches stays regular at $\epsilon \rightarrow 0$: namely,

\begin{align}
J^{({\rm reg})}_{2|4}(\epsilon) = \left( 6^{-1/4} \dfrac{\Gamma(3/2)}{\Gamma(7/12)\Gamma(11/12)} \right) J^{(1)}_{2|4}(\epsilon) - \left( 6^{+1/4} \dfrac{\Gamma(1/2)}{\Gamma(1/12)\Gamma(5/12)} \right) J^{(2)}_{2|4}(\epsilon)
\end{align}
\smallskip\\
Therefore, it is this branch that should be identified with the bounded area (regularity of the bounded area in the limit $\epsilon \rightarrow 0$ is fairly obvious). Comparing the bounded area with the regular branch of the integral discriminant, we find a hypergeometric identity

\begin{align}
V(\epsilon) = {\rm const} \cdot J^{({\rm reg})}_{2|4}(\epsilon)
\label{Identity}
\end{align}
\smallskip\\
Eq. (\ref{Identity}) is indeed a valid hypergeometric identity, which can be seen by expanding both sides in powers of $\epsilon$:

\begin{align}
V(\epsilon) = \pi - \dfrac{1}{16} \pi \epsilon + \dfrac{9}{1024} \pi \epsilon^2 - \dfrac{25}{16384} \pi \epsilon^3  + \dfrac{1225}{4194304} \pi \epsilon^4  + \ldots =
\end{align}

\begin{align}
J^{({\rm reg})}_{2|4}(\epsilon) = \dfrac{1}{4} - \dfrac{1}{64} \epsilon + \dfrac{9}{4096} \epsilon^2 - \dfrac{25}{65536} \epsilon^3 + \dfrac{1225}{16777216} \epsilon^4 + \ldots
\end{align}
\smallskip\\
what fixes ${\rm const} = 4 \pi$ in (\ref{Identity}). As one can see, the bounded area and corresponding integral discriminant indeed coincide, up to a constant factor.

The above calculation with invariants is certainly not the most practical way to evaluate the integral discriminant: it is rather a conceptual demonstration of equivalence between these quantities, emphasizing the importance of correct choice of the regular branch. Usually, a much more convenient way to do is to perform a series expansion directly in the integral (\ref{Intd}). Doing so automatically guarantees that a correct branch is chosen, and often allows to calculate the bounded volume in a very convenient way. As an illustration, consider the algebraic hypersurface

\begin{align}
a (x_1^2 + x_2^2 + x_3^2)^2 + b (x_1^2 + x_2^2 + x_3^2) x_4^2 + c x_4^4 = \hbar
\end{align}
\smallskip\\
in four-dimensional space. Using the relation with integral discriminants, we find the bounded volume

\begin{align}
V(a,b,c) = \int\limits_{-\infty}^{+\infty} dx_1 \int\limits_{-\infty}^{+\infty} dx_2 \int\limits_{-\infty}^{+\infty} dx_3 \int\limits_{-\infty}^{+\infty} dx_4 \ \exp\left( - \dfrac{a}{\hbar} (x_1^2 + x_2^2 + x_3^2)^2 - \dfrac{b}{\hbar} (x_1^2 + x_2^2 + x_3^2) x_4^2 - \dfrac{c}{\hbar} x_4^4 \right)
\end{align}
\smallskip\\
Passing to new integration variables

\[
\left\{
\begin{array}{l}
x_1 = \rho \cos(\theta_1) \\
\\
x_2 = \rho \sin(\theta_1) \cos(\phi) \\
\\
x_3 = \rho \sin(\theta_1) \sin(\phi) \\
\\
x_4 = t \\
\end{array}
\right.
\]
\smallskip\\
with

\begin{align}
0 \leq \theta \leq \pi, 0 \leq \phi \leq 2 \pi, 0 \leq \rho \leq \infty, -\infty \leq t \leq \infty
\end{align}
\smallskip\\
and Jacobian $\rho^2 \sin(\theta)$, we obtain

\begin{align}
V(a,b,c) = \int\limits_{0}^{2 \pi} d \phi \int\limits_{0}^{\pi} \sin(\theta) d \theta \int\limits_{0}^{\infty} \rho^2 d \rho \int\limits_{-\infty}^{\infty} d t \exp\left( - \dfrac{a}{\hbar} \rho^4 - \dfrac{b}{\hbar} \rho^2 t^2 - \dfrac{c}{\hbar} t^4 \right)
\end{align}
\smallskip\\
The angular integration is easily done, and we find

\begin{align}
V(a,b,c) = 4 \pi \int\limits_{0}^{\infty} \rho^2 d \rho \int\limits_{-\infty}^{\infty} d t \exp\left( - \dfrac{a}{\hbar} \rho^4 - \dfrac{b}{\hbar} \rho^2 t^2 - \dfrac{c}{\hbar} t^4 \right)
\end{align}
\smallskip\\
Expanding into series

\begin{align}
V(a,b,c) = 4 \pi \sum\limits_{k = 0}^{\infty} \dfrac{b^k}{k! \hbar^k} \int\limits_{0}^{\infty} \rho^{2 + 2k} \exp\left( - \dfrac{a}{\hbar} \rho^4 \right) d \rho \int\limits_{-\infty}^{\infty} t^{2k} \exp\left( - \dfrac{c}{\hbar} t^4 \right) d t
\end{align}
\smallskip\\
and evaluating both integrals, we find

\begin{align}
V(a,b,c) = \dfrac{\pi}{2} \sum\limits_{k = 0}^{\infty} \dfrac{b^k}{k! \hbar^k} \Gamma\left( \dfrac{k}{2} + \dfrac{1}{4} \right) \Gamma\left( \dfrac{k}{2} + \dfrac{3}{4} \right) \left( \dfrac{a}{\hbar} \right)^{-3/4 - k/2} \left( \dfrac{c}{\hbar} \right)^{-1/4 - k/2}
\end{align}
\smallskip\\
The sums are easily evaluated in elementary functions, and we finally obtain

\begin{align}
V(a,b,c) = \dfrac{\pi^2}{\sqrt{a} } \dfrac{\hbar}{\sqrt{b + 2\sqrt{ac}}}
\end{align}
\smallskip\\
In this example, the use of integral discriminant allows to obtain a simple formula for the volume in radicals. Thus integral discriminants are useful in application to the calculation of bounded volume.

\subsection{Algebraic numbers as hypergeometric series}

A characteristic property of integral discriminants, which is evident from definition, is $SL(n)$ invariance: integral

\begin{align*}
J_{n|r}\big(S\big) = \int dx_1 \ldots dx_n \ e^{-S(x_1, \ldots, x_n)}
\end{align*}
\smallskip\\
does not depend on the choice of basis in the $n$-dimensional linear space and, consequently, depends only on $SL(n)$ invariants $I_k(S)$ (this dependence is made explicit in the above summary table). This property is not generic: many other quantities, which belong to the class of partition functions, are not $SL(n)$ invariant.

An important example is provided by roots of polynomials, also known as algebraic numbers:

\begin{align*}
f_0 + f_1 \lambda + \ldots + f_r \lambda^r = 0
\end{align*}
\smallskip\\
This equation defines $\lambda$ as a function of $f_0, \ldots, f_r$:

\begin{align*}
\lambda = \lambda(f_0, \ldots, f_r)
\end{align*}
\smallskip\\
and this function is naturally multi-valued -- there are exactly $r$ branches of $\lambda$, since algebraic equation of degree $r$ has exactly $r$ roots in the complex plane. For low $r$ this function can be expressed in radicals. For $r = 2$ this is merely the solution of quadratic equation:

\begin{align*}
\lambda(f_0, f_1, f_2) = - \dfrac{1}{2} \dfrac{f_{1} \pm \sqrt{ f_{1}^2-4 f_{2} f_{0} }}{f_{2}}
 \end{align*}
\smallskip\\
Similarly, for $r = 3$ one of the roots is given by

\begin{align*}
\lambda(f_0, f_1, f_2, f_3) \ = \ & -\dfrac{1}{3} \dfrac{f_2}{f_3} + \dfrac{(36 f_1 f_2 f_3 - 108 f_0 f_3^2 - 8 f_2^3 + 12 f_3 \sqrt{12 f_{1}^3 f_{3}-3 f_{1}^2 f_{2}^2-54 f_{1} f_{2} f_{3} f_{0}+81 f_{0}^2 f_{3}^2+12 f_{0} f_{2}^3} )^{1/3}}{6f_3} - \emph{} \\ & \\ & \emph{} - \dfrac{6 f_1 f_3 - 2 f_2^2}{3f_3(36 f_1 f_2 f_3 - 108 f_0 f_3^2 - 8 f_2^3 + 12 f_3 \sqrt{12 f_{1}^3 f_{3}-3 f_{1}^2 f_{2}^2-54 f_{1} f_{2} f_{3} f_{0}+81 f_{0}^2 f_{3}^2+12 f_{0} f_{2}^3} )^{1/3}}
 \end{align*}
\smallskip\\
and the others are analogously expressed through cubic roots. For $r = 4$, radicals of order $1/4$ are already necessary, and for generic $r$ expression in terms of radicals is impossible (as rigorously shown by Abel and Galois). Still, despite it is impossible to express them through elementary functions, for any $r$ the algebraic numbers satisfy the same differential equations

\begin{equation}
\addtolength{\fboxsep}{5pt}
\boxed{
\begin{gathered}
\left( \dfrac{\partial}{\partial f_a} \dfrac{\partial}{\partial f_b} - \dfrac{\partial}{\partial f_c} \dfrac{\partial}{\partial f_d} \right) \lambda(f_0, \ldots, f_r) = 0
\end{gathered}
}\label{WardRoots}
\end{equation}
\smallskip\\
as integral discriminants do. For example, it is possible to check the Ward identities directly for $r = 2$:

\begin{align*}
\left( \dfrac{\partial}{\partial f_0} \dfrac{\partial}{\partial f_2} - \dfrac{\partial}{\partial f_1} \dfrac{\partial}{\partial f_1} \right) \left( \dfrac{f_{1} \pm \sqrt{ f_{1}^2-4 f_{2} f_{0} }}{f_{2}} \right) = 0
\end{align*}
\smallskip\\
similarly for $r = 3$

\begin{align*}
\left( \dfrac{\partial}{\partial f_0} \dfrac{\partial}{\partial f_2} - \dfrac{\partial}{\partial f_1} \dfrac{\partial}{\partial f_1} \right) & \left(-\dfrac{1}{3} \dfrac{f_2}{f_3} + \dfrac{(36 f_1 f_2 f_3 - 108 f_0 f_3^2 - 8 f_2^3 + 12 f_3 \sqrt{12 f_{1}^3 f_{3}-3 f_{1}^2 f_{2}^2-54 f_{1} f_{2} f_{3} f_{0}+81 f_{0}^2 f_{3}^2+12 f_{0} f_{2}^3} )^{1/3}}{6f_3} - \emph{} \right.
 \end{align*}

\begin{align*}
\left. \emph{} - \dfrac{6 f_1 f_3 - 2 f_2^2}{3f_3(36 f_1 f_2 f_3 - 108 f_0 f_3^2 - 8 f_2^3 + 12 f_3 \sqrt{12 f_{1}^3 f_{3}-3 f_{1}^2 f_{2}^2-54 f_{1} f_{2} f_{3} f_{0}+81 f_{0}^2 f_{3}^2+12 f_{0} f_{2}^3} )^{1/3}}\right) = 0
 \end{align*}

\begin{align*}
\left( \dfrac{\partial}{\partial f_0} \dfrac{\partial}{\partial f_3} - \dfrac{\partial}{\partial f_1} \dfrac{\partial}{\partial f_2} \right) & \left(-\dfrac{1}{3} \dfrac{f_2}{f_3} + \dfrac{(36 f_1 f_2 f_3 - 108 f_0 f_3^2 - 8 f_2^3 + 12 f_3 \sqrt{12 f_{1}^3 f_{3}-3 f_{1}^2 f_{2}^2-54 f_{1} f_{2} f_{3} f_{0}+81 f_{0}^2 f_{3}^2+12 f_{0} f_{2}^3} )^{1/3}}{6f_3} - \emph{} \right.
 \end{align*}

\begin{align*}
\left. \emph{} - \dfrac{6 f_1 f_3 - 2 f_2^2}{3f_3(36 f_1 f_2 f_3 - 108 f_0 f_3^2 - 8 f_2^3 + 12 f_3 \sqrt{12 f_{1}^3 f_{3}-3 f_{1}^2 f_{2}^2-54 f_{1} f_{2} f_{3} f_{0}+81 f_{0}^2 f_{3}^2+12 f_{0} f_{2}^3} )^{1/3}}\right) = 0
 \end{align*}\smallskip\\
and also for $r = 4$. For arbitrary $r$ (when no expressions through radicals are available) validity of Ward identities (\ref{WardRoots}) becomes obvious, if one recalls the conventional Cauchy integral representation for $\lambda$:

\begin{align}
\lambda(f_0, \ldots, f_r) = \oint z \ d \log \Big( f_0 + f_1 z + \ldots + f_r z^r \Big) = - \oint \log \Big( f_0 + f_1 z + \ldots + f_r z^r \Big) \ d z
\end{align}
\smallskip\\
All this -- multivaluedness, Ward identities and integral representations -- make algebraic numbers quite similar to integral discriminants. Of course, since roots are not $SL$-invariant, they are not literally equal to integral discriminants. Still, as a consequence of Ward identities, solutions of algebraic equations are given by hypergeometric functions, similarly to integral discriminants. The basic example here is the case of $r = 2$:

\begin{align}
f_0 + f_1 \lambda + f_2 \lambda^2 = 0
\end{align}
\smallskip\\
This equation has two independent solutions, given by explicit formulas

\begin{align}
\lambda_{+}(f_0, f_1, f_2) = - \dfrac{1}{2} \dfrac{f_{1} + \sqrt{ f_{1}^2-4 f_{2} f_{0} }}{f_{2}}, \ \ \ \ \
\lambda_{-}(f_0, f_1, f_2) = - \dfrac{1}{2} \dfrac{f_{1} - \sqrt{ f_{1}^2-4 f_{2} f_{0} }}{f_{2}}
\end{align}
\smallskip\\
If viewed as a series, these functions take a form

\begin{align}
x_1 = - \dfrac{f_0}{f_1} - \dfrac{f_0^2 f_2}{f_1^3} - \dfrac{2 f_0^3 f_2^2}{f_1^5} - \dfrac{5 f_0^4 f_2^3}{f_1^7} - \dfrac{14 f_0^5 f_2^4}{f_1^9} - \dfrac{42 f_0^6 f_2^5}{f_1^{11}} - \ldots
\end{align}
\smallskip\\
and

\begin{align}
x_2 = - \dfrac{f_1}{f_2} + \dfrac{f_0}{f_1} + \dfrac{f_0^2 f_2}{f_1^3} + \dfrac{2 f_0^3 f_2^2}{f_1^5} + \dfrac{5 f_0^4 f_2^3}{f_1^7} + \dfrac{14 f_0^5 f_2^4}{f_1^9} + \dfrac{42 f_0^6 f_2^5}{f_1^{11}} + \ldots
\end{align}
\smallskip\\
Given these series, one can also verify the validity of Ward identities term-by-term. Note again the similarity with integral discriminants: there are two branches, which satisfy one and the same differential equation and differ only by asymptotics (at large $f_1$, one branch grows asymptotically as $-f_1/f_2$, while the other tends to zero). The coefficients here are famous Catalan numbers,

\begin{align}
1, 2, 5, 14, 42, \ldots = \dfrac{(2k)!}{k!(k+1)!}
\end{align}
\smallskip\\
They are integers and have a celebrated combinatorial meaning of counting the number of tree diagrams, which appear in the Bogolubov-like iteration procedure \cite{Bogolubov}. However, we are interested not so much in diagrammatical and/or combinatorial meaning of these numbers, but rather in their hypergeometric nature: the solutions are given by explicit hypergeometric series

\begin{align}
x_1 = - \sum\limits_{k = 0}^{\infty} \dfrac{(2k)!}{k!(k+1)!} \dfrac{f_0^{k+1} f_2^{k}}{f_1^{2k+1}} = - \dfrac{f_0}{f_1} \cdot \ _{2}F_{1} \left( \left[ \dfrac{1}{2}, 1 \right], \left[ 2 \right], \dfrac{4 f_2 f_0}{f_1^2} \right)
\end{align}
and

\begin{align}
x_1 = - \dfrac{f_1}{f_2} + \sum\limits_{k = 0}^{\infty} \dfrac{(2k)!}{k!(k+1)!} \dfrac{f_0^{k+1} f_2^{k}}{f_1^{2k+1}} = - \dfrac{f_1}{f_2} + \dfrac{f_0}{f_1} \cdot \ _{2}F_{1} \left( \left[ \dfrac{1}{2}, 1 \right], \left[ 2 \right], \dfrac{4 f_2 f_0}{f_1^2} \right)
\end{align}
\smallskip\\
which just occasionally reduce to square roots. Note again the similarity with integral discriminants: singularity of these series is situated at the unit value of argument, when $f_1^2 = 4 f_2 f_0$, i.e, when the algebraic discriminant vanishes. In variance with expressions through radicals, expressions through hypergeometric functions can be written for arbitrary $r$: say, an equation $x^6 + x + c = 0$ which is not solvable in radicals, can be solved in series

\begin{align}
x = - c - c^6 - 6 c^{11} - 51 c^{16} - 506 c^{21} - 5481 c^{26} - 62832 c^{31} - 749398 c^{36} + \ldots
\end{align}
\smallskip\\
Looking at these numbers, it is easy to find

\begin{align}
1,6,51,506,5481,62832,749398, \ldots = \dfrac{(6k)!}{k!(5k+1)!}
\end{align}
\smallskip\\
These numbers are again integer and enumerate sextic (6-ary) trees (rooted, ordered, incomplete) with $k$ vertices (including the root). Moreover, their generating function is again hypergeometric:

\begin{align}
x = - \sum\limits_{k = 0}^{\infty} \dfrac{(6k)!}{k!(5k+1)!} c^{k+1} = - c \cdot \ _{5}F_{4}\left( \left[\dfrac{1}{6},\dfrac{2}{6},\dfrac{3}{6},\dfrac{4}{6},\dfrac{5}{6} \right], \left[ \dfrac{2}{5},\dfrac{3}{5},\dfrac{4}{5} \right], \dfrac{46656c}{3125} \right)
\end{align}
\smallskip\\
Note, that this function represents only one of the roots: the other roots are represented by other hypergeometric functions, which are completely analogous and which we do not include here. For more details about hypergeometricity of solutions of algebraic equations, see \cite{Sturmfels}. Note also, that integerness of the coefficients is not a generic property, it is due to particular form of the above examples. Say, algebraic equation

$$ (x^2 + a)^3 + (x + b)^4 = 0 $$
\smallskip\\
has a solution

$$x = I + \dfrac{3I}{2} a + 2 b - \dfrac{21I}{8} a^2 - 6 a b + 3I b^2 + \ldots $$
\smallskip\\
with already non-integer coefficients and exact formula

$$x = I + \dfrac{1}{2} \sum\limits_{s,m = 0}^{\infty}  (-1)^{s + 1} (2I)^{m+1} \dfrac{\Gamma\left( 3s + \dfrac{3}{2}m - \dfrac{1}{2} \right)\Gamma\left( 2s + \dfrac{3}{2}m - 1 \right)}{\Gamma\left( 2s + m - \dfrac{1}{2} \right)\Gamma\left( 2s + m \right)} \dfrac{a^{s}}{s!} \dfrac{b^m}{m!} $$
\smallskip\\

Let us make one more short note about $SL(n)$ invariance. As already mentioned, the roots theirselves are not $SL(2)$ invariant: under transformation $G \in SL(2)$ they transform as

\begin{align}
\lambda \mapsto \dfrac{G_{11} \lambda + G_{12}}{G_{21} \lambda + G_{22}}, \ \ \ \ \ \ G_{11} G_{22} - G_{12} G_{21} = 1
\end{align}
\smallskip\\
Thus they are not expressible through the elementary invariants. However, the double ratios

\begin{align}
\Lambda_{ijkl} = \dfrac{(\lambda_i - \lambda_j) (\lambda_k - \lambda_l)}{(\lambda_i - \lambda_k) (\lambda_j - \lambda_l)}
\end{align}
\smallskip\\
are invariant under $SL(n)$ transformations (this is easy to check directly). For example, for $r = 4$

\begin{align}
f_0 + f_1 \lambda + f_2 \lambda^2 + f_3 \lambda^3 + f_4 \lambda^4 = 0
\end{align}
\smallskip\\
there are two elementary invariants

\begin{align}
I_{2} = 2 f_{0} f_{4} - \dfrac{1}{2} f_{1} f_{3} + \dfrac{1}{6} f_{2}^2
\end{align}

\begin{align}
I_{3} = f_{0} f_{2} f_{4}- \dfrac{3}{8} f_{0} f_{3}^2- \dfrac{3}{8} f_{1}^2 f_{4}+ \dfrac{1}{4} f_{1} f_{2} f_{3}- \dfrac{1}{36} f_{2}^3
\end{align}
\smallskip\\
and a single double ratio $\Lambda_{1234}$, which is $SL(2)$ invariant and thus expressible through $I_2$ and $I_3$. To express $\Lambda_{1234}$ through the elementary invariants, let us consider a particular case

\begin{align}
f_0 = -A^4, \ f_1 = \epsilon, \ f_2 = 0, \ f_3 = 0, \ f_4 = 1
\label{SpecificAe}
\end{align}
\smallskip\\
For these values of parameters $I_2 = -2 A^4$ and $I_3 = - 3 \epsilon^2/8$. It is easy to obtain, that

\[
\left\{
\begin{array}{lllll}
\lambda_1 = A - \dfrac{1}{4} \dfrac{\epsilon}{A^2} -\dfrac{1}{32} \dfrac{\epsilon^2}{A^5} + \dfrac{7}{2048}\dfrac{\epsilon^4}{A^{11}} +\dfrac{1}{512} \dfrac{\epsilon^5}{A^{14}} + \ldots \\
\\
\lambda_2 = A I+\dfrac{1}{4}\dfrac{\epsilon}{A^2}+\dfrac{1}{32} I\dfrac{\epsilon^2}{A^5}+\dfrac{7}{2048} I\dfrac{\epsilon^4}{A^{11}}-\dfrac{1}{512}\dfrac{\epsilon^5}{A^{14}} + \ldots \\
\\
\lambda_3 = -A-\dfrac{1}{4}\dfrac{\epsilon}{A^2}+\dfrac{1}{32}\dfrac{\epsilon^2}{A^5}-\dfrac{7}{2048}\dfrac{\epsilon^4}{A^{11}}+\dfrac{1}{512}\dfrac{\epsilon^5}{A^{14}} + \ldots \\
\\
\lambda_4 = -I A+\dfrac{1}{4}\dfrac{\epsilon}{A^2}-\dfrac{1}{32} I\dfrac{\epsilon^2}{A^5}-\dfrac{7}{2048} I\dfrac{\epsilon^4}{A^{11}}-\dfrac{1}{512}\dfrac{\epsilon^5}{A^{14}} + \ldots \\
\end{array}
\right.
\]
\smallskip\\
These four branches correspond to $\epsilon$-deformations of the four roots of the simple equation $\lambda^4 = A^4$.
Accordingly, the double ratio can be calculated as a series in $\epsilon$ and equals

\begin{align}
\Lambda_{1234} = \dfrac{(\lambda_1 - \lambda_2) (\lambda_3 - \lambda_4)}{(\lambda_1 - \lambda_3) (\lambda_2 - \lambda_4)} = \dfrac{1}{2} - \dfrac{3I}{32} \dfrac{\epsilon^2}{A^6} + \dfrac{33I}{16384} \dfrac{\epsilon^6}{A^{18}} - \dfrac{1665I}{16777216} \dfrac{\epsilon^{10}}{A^{30}} + \ldots
\end{align}
\smallskip\\
Expressed through invariants, it takes form

\begin{align}
\Lambda_{1234} = \dfrac{1}{2} - \dfrac{\sqrt{2}}{2} \dfrac{I_3}{I_2^{3/2}} - \dfrac{11\sqrt{2}}{18} \dfrac{I_3^3}{I_2^{9/2}} - \dfrac{185\sqrt{2}}{108} \dfrac{I_3^5}{I_2^{{15}/2}} - \ldots
\end{align}
\smallskip\\
This is the simplest way to express the double ratio through $I_2$ and $I_3$. Because of its $SL(2)$ invariance, this equality will remain the same for arbitrary $f_0, \ldots, f_4$, not just for particular values (\ref{SpecificAe}).

Let us finish this section by noting, that all algebraic numbers -- in other words, the set of numbers which are solutions of \emph{some} algebraic equations -- form a field. This in particular means that the set of all algebraic numbers is closed under addition and multiplication. For example, suppose that $x_1$ and $x_2$ are algebraic numbers. I.e, suppose that $x_1$ is a root of $f(z)$ and $x_2$ is a root of $g(z)$. Then polynomials $f(z)$ and $g(t - z)$ have a common root for $t = x_1 + x_2$, so that $x_1 + x_2$ is a root of resultant of $f(z)$ and $g(t - z)$. Since resultant is a polynomial, $x_1 + x_2$ is an algebraic number. Similarly one can prove that $x_1 - x_2$, $x_1 x_2$ and $x_1 / x_2$ for $x_2 \neq 0$ are algebraic numbers. Thus, algebraic numbers actually form a field.

By analogy, one could in principle consider 2-algebraic numbers $(x,y)$: solutions of systems of equations

\[
\left\{
\begin{array}{ccc}
f(x,y) = 0,\\
\\
g(x,y) = 0
\end{array}
\right.
\]
\smallskip\\
where $f$ and $g$ are some non-homogeneous polynomials, and their straightforward generalisations for arbitrary number $n$ of independent variables. However, multiplication and division of such multi-component objects seems to be unnatural. A more natural direction of generalisation makes use of the similarity between algebraic numbers and integral discriminants: it is interesting to understand, in what exact sence the set of integral discriminants of homogeneous forms is a ring or even a field. This brings us directly into the field of motivic theories (see \cite{Motivic} for a recent review "for physicists") which is, however, beyond the scope of this paper.

\section{Conclusion}

We have briefly described (hopefully in explicit and elementary way) several research lines in resultant theory. These include Sylvester-Bezout (determinantal) and Schur (analytic) formulas for the resultant, invariant formulas and symmetry reductions for the discriminant, Ward identities and hypergeometric series representations for partition functions (integral discriminants and roots of algebraic equations). Unfortunately, several important directions -- such as Macaulay matrices \cite{Macaulay}, Newton polytopes \cite{GKZ}, Mandelbrot sets \cite{Mandelbrot} and others -- remain untouched in this review. They will be discussed elsewhere.

Let us emphasise that the objects and relations, which we consider in present paper, are somewhat under-investigated. Many interesting properties of resultants, discriminants and especially of partition functions have attracted attention only recently. Moreover, new unexpected properties and relations continue to appear. All this indicates, that resultant theory and the entire non-linear algebra is only in its beginning.

\section*{Acknowledgements}

Our work is partly supported by Russian Federal Nuclear Energy Agency, Russian Federal Agency for Science and Innovations under contract 02.740.11.5029 and the Russian President's Grant of Support for the Scientific Schools NSh-3035.2008.2, by RFBR grant 07-02-00645, by the joint grants 09-02-90493-Ukr, 09-01-92440-CE, 09-02-91005-ANF and 09-02-93105-CNRS. The work of Sh.Shakirov is also supported in part by the Moebius Contest Foundation for Young Scientists.

\end{document}